\documentclass[pdflatex,sn-nature]{sn-jnl}

\usepackage[utf8]{inputenc}

\usepackage{graphicx}%
\usepackage{multirow}%
\usepackage{amsmath,amssymb,amsfonts}%
\usepackage{amsthm}%
\usepackage[cal=rsfs]{mathalfa}

\usepackage[title]{appendix}%
\usepackage{xcolor}%
\usepackage{textcomp}%
\usepackage{manyfoot}%
\usepackage{booktabs}%
\usepackage{algorithm}%
\usepackage{algorithmicx}%
\usepackage{algpseudocode}%
\usepackage{listings}%
\usepackage[normalem]{ulem}
\usepackage{bm}

\theoremstyle{thmstyleone}%
\theoremstyle{thmstyletwo}%

\theoremstyle{thmstylethree}%

\usepackage{subcaption}
\usepackage{tikz}

\DeclareMathOperator{\sign}{sign} 
\DeclareMathOperator{\Ai}{Ai} 
\DeclareMathOperator{\Bi}{Bi} 
\DeclareMathOperator{\cn}{cn}
\DeclareMathOperator{\am}{am}
\newcommand{\D}[0]{{\rm d}}

\usepackage{geometry}
\usepackage{caption}
\DeclareCaptionLabelFormat{none-but-active}{} 
\newcommand{\figuresection}[1]{%
    \subsection*{#1}
    \addcontentsline{toc}{subsection}{#1}
}

\begin{document}

\title[Article Title]{Shaping nematic order in bacterial films with single-cell resolution patterning}


\author[1]{\fnm{Matthias} \sur{Le Bec}}

\author[2]{\fnm{Guillem} \sur{Pérez Martín}}

\author[1]{\fnm{Cameron} \sur{Boggon}}

\author[1]{\fnm{Yiyao} \sur{Hu}}

\author[2]{\fnm{Leonardo} \sur{Puggioni}}

\author[4]{\fnm{Rosa} \sur{Heydenreich}}

\author[4]{\fnm{Alexander} \sur{Mathys}}

\author*[2]{\fnm{Luca} \sur{Giomi}}\email{giomi@lorentz.leidenuniv.nl}

\author*[3]{\fnm{Eleonora} \sur{Secchi}}\email{secchi@ifu.baug.ethz.ch}

\author*[1]{\fnm{Lucio} \sur{Isa}}\email{lucio.isa@mat.ethz.ch}

\affil[1]{\orgdiv{Materials Department}, \orgname{ETH Zürich}, \orgaddress{\street{Vladimir-Prelog-Weg 1-5/10}, \city{Zürich}, \postcode{8093}, \country{Switzerland}}}

\affil[2]{\orgdiv{
Instituut-Lorentz}, \orgname{Universiteit Leiden}, \orgaddress{\street{P.O. Box 9506}, \city{Leiden}, \postcode{ 2300 RA}, \country{The Netherlands}}}

\affil[3]{\orgdiv{Civil, environmental and Geomatic Engineering Department}, \orgname{ETH Zürich}, \orgaddress{\street{Laura-Hezner-Weg 7}, \city{Zürich}, \postcode{8093}, \country{Switzerland}}}

\affil[4]{\orgdiv{Health Sciences and Technology Department}, \orgname{ETH Zürich}, \orgaddress{\street{Universitätstrasse 2}, \city{Zürich}, \postcode{8092}, \country{Switzerland}}}


\abstract{
Bacterial colonies composed of elongated cells form active nematic fluids that spontaneously self-organise into ordered domains of aligned cells and exhibit self-generated chaotic flows powered by cell growth. While their dynamics have attracted significant attention, the role of initial conditions remains largely unexplored due to a lack of precise patterning methods. Here, we harness the precision of capillary assembly to pattern \textit{Bacillus subtilis} endospores into arrays with controlled positions and orientations at single-cell resolution. Upon germination and growth of cell chains, we quantify the dynamics and morphologies of the resulting bacterial films. While orthogonally seeded spores lead to chaotic dynamics, seeding them with parallel orientations yields films with high nematic order across millimetres, which subsequently synchronously buckle upon further growth. Our observations are captured by numerical simulations and a model that describes the buckling dynamics starting from the mechanical properties of individual filaments. By programming local cell orientation with single-cell precision, we finally harness nematic alignment to create macroscopic bacterial films with local optical anisotropy, via structural colouration and light polarisation. Our findings demonstrate that initial conditions play a key role and offer exciting opportunities to control the spatio-temporal organization of bacterial assemblies towards addressing open biological questions and realizing living materials with tailored properties.
}




\maketitle

Bacteria that come into contact with surfaces often form biofilms, multicellular assemblies embedded in an extracellular matrix that provides structural integrity and protection against external stresses~\cite{sauer_biofilm_2022}. It is now established that the spatial organisation of bacterial cells within biofilms determines their ecological behaviour and response to external stimuli ~\cite{nadell_spatial_2016, hallatschek_proliferating_2023}. Cell orientation, aspect ratio and cell-cell alignment influence biofilm mechanical properties and their response to antibiotic treatment~\cite{diaz-pascual_breakdown_2019,hartmann_emergence_2019,jeckel_shared_2022,nijjer_biofilms_2023}. Tight cell packing has, for instance, been linked with higher biofilm resistance to external stresses~\cite{zhang_community-specific_2023,nahum_effect_2024} and limits invasion by exogenous microorganisms~\cite{diaz-pascual_breakdown_2019}. However, these properties have largely been characterised in biofilms that emerge from uncontrolled growth, and the ability to predict and actively control spatial organisation—and its functional consequences—remains limited. 

Conversely, the paradigm of {\em active nematics}~\cite{doostmohammadi_active_2018, doostmohammadi_physics_2022} provides an efficient conceptual physical framework to decipher various structural and dynamical properties of assemblies of rod-shaped bacteria. Whether free-swimming (i.e. planktonic) or surface-associated (i.e. sessile), individual bacteria generate extensile active stresses that lead to the emergence of structures across distances far exceeding the typical cellular size. 
Some communities of planktonic bacteria for instance display a distinctive swarming behaviour mirroring 
a hallmark active nematic phenomenon known as ``active turbulence''~\cite{secchi_intermittent_2016,li_data-driven_2019,yashunsky_topological_2024}. By contrast, monolayers of sessile bacteria proliferating on a solid substrate organise in clusters of highly aligned cells~\cite{you_geometry_2018,dellarciprete_growing_2018}, whose distribution determines the evolution of a colony under confinement~\cite{volfson_biomechanical_2008,you_confinement-induced_2021} and into multi-layered structures~\cite{beroz_verticalization_2018, you_mono-_2019}. Furthermore, growth-induced active stresses generate vortical flows that enhance cell mixing and thus promote a high genetic diversity within growing colonies~\cite{schwarzendahl_active_2022}. 

Among the features of active nematic systems, physical mechanisms based on topological defects — particularly $\pm 1/2$ disclinations — have attracted considerable interest due to their robustness and reproducibility. In two-dimensional nematic liquid crystals, including bacterial monolayers, disclinations appear as singular points around which the orientation field $\theta$ rotates by $180^{\circ}$ increments, such that $\oint \mathrm{d}\theta = 2\pi s$ with $s=\pm 1/2,,\pm 3/2,\ldots$. This sharp reorientation focuses active stresses, generating spatially extended and yet localised regions of large shear that colonies can harness to achieve mechano-biological functionality. These include the regulation of global migration speed in crowded environments~\cite{meacock_bacteria_2021}, the formation of protrusions along the colony's boundary~\cite{doostmohammadi_defect-mediated_2016} and the selection of specific bacterial phenotypes during the invasion of empty spaces~\cite{basaran_large-scale_2022}. 
Yet, despite their prominent regulatory role in bacterial colonies, the link between initial cell patterning and the emergence and evolution of topological defects during growth remains poorly understood. In particular, it is unclear how defect structures arise during proliferation, how they shape the early-time evolution of the colony, and how colony growth in turn feeds back on defect density and spatial organisation. This gap limits a mechanistic understanding of the role of nematic order in the development of bacterial communities. A key challenge is that surface-attached communities typically grow from randomly seeded cells, making it difficult to control where defects emerge, as well as their number and temporal evolution. 


To shed light onto these phenomena, here we introduce an experimental approach that allows us to achieve spatio-temporal control on topological defects in monolayers of \textit{Bacillus subtilis} cells. We use Capillarity-Assisted Particle Assembly (CAPA), to seed \textit{B. subtilis} endospores into 2D arrays with controlled positions and orientations (Figure \ref{fig_CAPAspore}A). Endospores, hereafter simply called spores, are bacterial cells in a dormant state, which some species belonging to the phylum Bacillota (formerly Firmicutes) -- including \textit{B. subtilis} -- revert to as a survival mechanism under harsh conditions. These spores can then germinate into vegetative cells upon nutrient exposure and grow as filaments of cells when cultivated on a solid surface~\cite{gestel_cell_2015}.  Following division, daughter cells remain attached, due to incomplete cell separation,  forming elongating chains or filaments of rod-shaped bacteria whose width corresponds to that of a single cell.  At high cell density, those filaments interact and can locally align with one another leading to the emergence of active nematic order. Combining experiments with numerical simulations and analytical work, we show that growing filaments can spontaneously buckle generating pairs of $\pm 1/2$ nematic disclinations, thereby imprinting the initial state of a two-dimensional active turbulent flow. Moreover, by engineering the spatial arrangement and orientation of the cells, we demonstrate that the early-time dynamics of the colony can be steered toward specific collective behaviours, including large-scale defect-free monolayers with high degree of nematic order, hierarchical collective buckling events or the complete loss of nematic order. By controlling the initial position and alignment of bacteria at the single-cell level, we take one step further toward achieving predictable microstructures in biofilms, which is an essential step for the development of novel, sustainable biomaterials with high-resolution control of local properties. As an example, we show here the growth of bacterial films with tailored anisotropic optical properties arising from  structural colouration originating from local nematic ordering. 

\newpage

\section*{Topological patterning of bacterial colonies via CAPA}\label{sec1}

Capillary assembly methods enable  the rational design of microparticle patterns with arbitrary 2D arrangement and single-particle resolution\cite{ni_capillary_2018}. CAPA utilises the capillary forces of a receding liquid-air meniscus to load colloids into microfabricated traps that match the dimensions of the particles (Figure S1). Driven by evaporation, colloids in suspension accumulate at the meniscus, which applies capillary forces that push them inside the traps and remove any non-trapped ones. After deposition, the substrate and the particles are then left to dry in air. While irrelevant for non-biological objects, both capillary forces and air drying impose high stress on microorganisms. 
Although certain bacteria, e.g. \textit{Staphylococcus aureus}, can withstand such stresses\cite{boggon_single-cell_2023}, the viability of other species, e.g. some \textit{E. coli} strains, is severely compromised by the patterning process\cite{pioli_patterning_2021}. 

\begin{figure}[h]
\centering
\makebox[\textwidth][c]{
    \includegraphics[width=\textwidth]{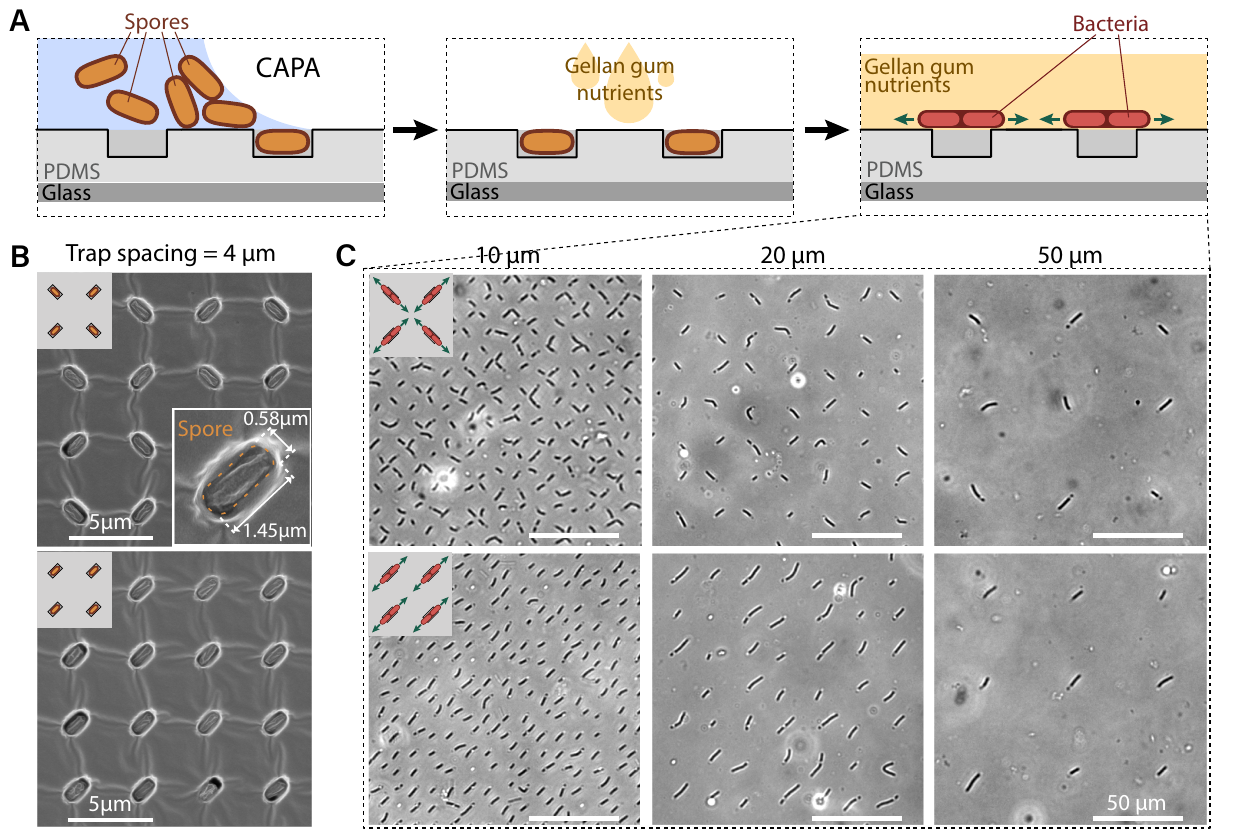}
    }
\caption{\textbf{Fig. 1 Single-cell control of bacteria position and growth orientation using capillarity-driven spore trapping.} (A) Scheme of the experimental procedure. Capillarity-Assisted Particle Assembly (CAPA) is performed with a \textit{B. subtilis} spore suspension over a PDMS surface micropatterned with traps of sizes fitting a single spore. Pouring melted nutritive gel directly onto the spore-loaded surface induces synchronised germination and bacterial outgrowth in 2D between the PDMS and the cured gel. (B) SEM images of spore-loaded PDMS surfaces for 4~$\mu$m spaced trap arrays with perpendicular (top) or parallel (bottom) orientations. Inset is a zoom-in on one spore in a trap, the dimensions of this spore are 1.45$\times$0.58 $\mu$m. (C) Bright-field images of bacterial growth out of the traps $\sim$2h30 after gel curing for arrays with 10, 20 and 50~$\mu$m spacing and with parallel (top) or perpendicular (bottom) orientations.
}\label{fig_CAPAspore}
\end{figure}

To circumvent these issues, we use metabolically inactive \textit{B. subtilis} spores, allowing deposition to be performed while the cells are resistant to dessication and effectively decouple the deposition process from bacterial growth. Moreover, \textit{B. subtilis} spores are anisotropic, with an aspect ratio of $\approx 2$ (i.e. dimensions of 1.45$\times$0.58~$\mu$m), enabling control over their initial orientation through the use of oriented rectangular traps  (Figure \ref{fig_CAPAspore}B). We find that an optimal trap size of 2$\times$1.1$\times$0.8~$\mu$m minimises the amount of empty and double-loaded traps and we then microfabricate a series of 2D arrays of these traps with varying spacing distances from 4 to 100~$\mu$m and with parallel or perpendicular orientations (Figure \ref{fig_CAPAspore}C). After filling the traps with spores using CAPA, spore-loaded templates can be stored desiccated for at least 10 days without loss of viability. We can then trigger synchronous spore germination into vegetative cells on demand by pouring a melted hydrogel containing nutrients (Gellan gum + Tryptone Soy Broth) onto the templates (Figure \ref{fig_CAPAspore}A and S2). The bacterial outgrowth is directed along the trap-imposed spore orientation (Figure \ref{fig_CAPAspore}A and C) and typically initiates from one end of the long axis of the trap. The choice of which end is stochastic, and this initial asymmetric growth leads to one end of bacterial filaments to be temporarily retained in the trap. After a few cell divisions, all bacteria ultimately grow out of the traps in both directions. The bacteria chains then grow exponentially along their long axis, while being confined in 2D between the PDMS substrate and the gel.

Our method achieves micron-scale resolution in the control of the initial spatial organisation of microbial communities, compared to other printing, self-assembly, or cell communication methods which remain limited to the $\sim$100~$\mu$m scale resolution~\cite{barbier_engineering_2022,herzog_3d_2024,xiao_bacterial_2024}

\section*{Defect unbinding and buckling in isolated bacterial chains}\label{sec2}

\begin{figure}[h]
\centering
\makebox[\textwidth][c]{
    \includegraphics[width=\textwidth]{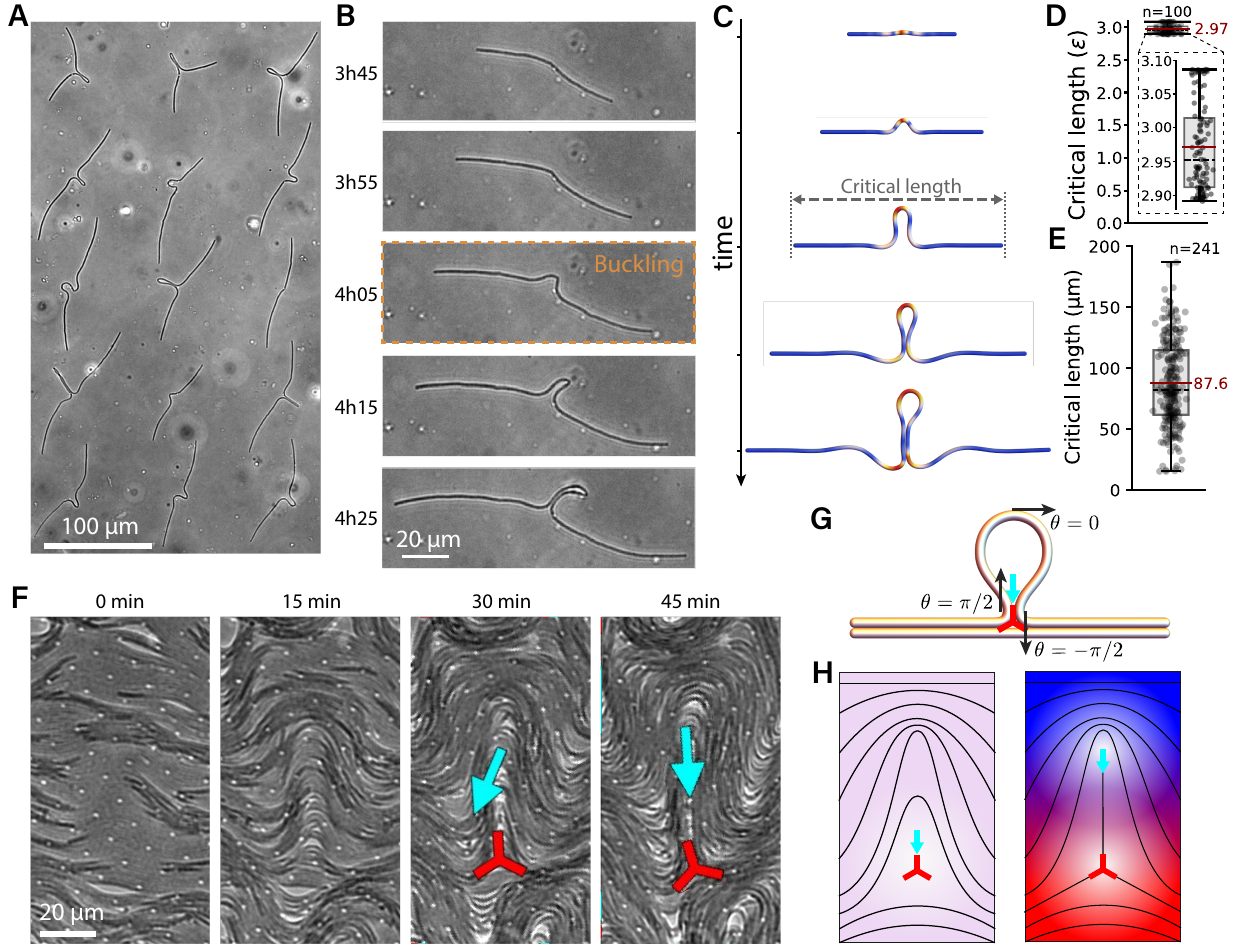}
    }
\caption{\textbf{Fig. 2 Isolated chain buckling and nematic defect creation.} (A) Bright-field image of isolated bacteria chains obtained by patterning spores in traps spaced by $100 \mu$m exemplifying different buckling modes. Experimental (B) and simulation (C) time-series of an isolated bacteria chain undergoing buckling. The colour scale corresponds to the local curvature, and thus the bending stress, with blue indicating mechanically relaxed regions and red indicating strongly loaded regions. Quantification of the numerical (D) and experimental (E) buckling critical length for isolated bacteria chains. The numerical data are rescaled by the theoretical length scale $\xi=(B/f)^{1/3}$. (F) Emergence and unbinding of a nematic defect dipole in a confluent bacterial film. (G-H) Schematic illustration of the same process shown in panel F. The colors highlight the transition from a topological neutral configuration (light purple) to a dipolar structure (red and blue).}
\label{fig_SingleBuckling}
\end{figure}

As individual bacterial chains grow out of the traps, the interplay between growth and friction with the substrate causes buckling of the chain around the middle, which leads to the unbinding of a pair of $\pm 1/2$ disclinations, with prescribed initial position and orientation, as illustrated in Figure~\ref{fig_SingleBuckling}.

We produce spore arrays with 100~$\mu$m spacing to obtain numerous isolated chains within each image and track the length at which they buckle (Figure \ref{fig_SingleBuckling}A, Supplementary Movie 1). After an initial phase of uniaxial growth along the trap orientation, chains buckle around their midpoints forming looped structures   (Figure \ref{fig_SingleBuckling}B). The loops close rapidly, with each side making contact and aligning with the other. 

A theoretical framework shows that   that buckling is driven by the cooperative effect of cell growth and drag caused by the frictional interactions with the substrate~\cite{yaman_emergence_2019, mcmahon_mechanical_2022, mcmahon_kinking_2025}. As the chains grow, they accumulate reactive forces at their tips, resulting in the build-up of compression along their length, which peaks at the midpoint. As in other buckling phenomena, the competition between compression and bending sets a length scale -- i.e. $\xi=(B/f)^{1/3}$, with $B$ the bending rigidity of the filaments and $f$ the force per unit length experienced by the filament at the tips. $\xi$ corresponds to the radius of curvature induced by the forcing, and, when it becomes comparable to the length of the filament, buckling occurs at the centre (Figure \ref{fig_SingleBuckling}C). Specifically, our analysis predicts a critical length approximatively three times larger than $\xi$: i.e. $L_{c}/\xi\approx 3$ (see SI Sec. $2$ for details). 

The analysis of the buckling instability is complemented by numerical simulations, where each filament is modelled as a semi-flexible bead-spring chain subjected to elastic compression, self-adhesion, friction and viscous drag and whose length grows exponentially in time (see SI Sec. $3$). In the absence of phenotypical noise -- that is, assuming uniform material properties -- our simulations yield a nearly constant critical length just below $3~\xi$ (Figure \ref{fig_SingleBuckling}D and SI Sec $3$).

We measured experimentally a critical buckling length of $\sim$88~$\mu$m (Figure \ref{fig_SingleBuckling}E), 


which provides a quantitative reference to map the simulation length scale to physical units, allowing the numerical results to be rescaled and directly compared with experiments. The measured value is significantly lower than its value obtained at the air-agarose interface ($\sim$140~$\mu$m) \cite{yaman_emergence_2019}, which can be explained by the fact that, in our case, cells must deform the nutritive gel, increasing the normal load and thus the friction forces. 
The experimental critical length also shows a large variance, reflecting the sensitivity to imperfection inherent in buckling~\cite{budiansky_theory_1974,li_mechanics_2012}, where small perturbations in filament properties or substrate heterogeneity lead to premature buckling.

When a filament is embedded in a dense colony, buckling drives the unbinding of a pair of $\pm 1/2$ nematic disclinations (Figure~\ref{fig_SingleBuckling}F). This mechanism is illustrated in Figure~\ref{fig_SingleBuckling}G,H. Consider, for simplicity, two initially straight filaments adhering to each other (Figure~\ref{fig_SingleBuckling}G). As filament buckles and starts forming a loop, its tangent vector rotates by $180^{\circ}$ before re-contacting the other filament. Adopting a bacterial active nematics convention, where the nematic director is parallel to the local tangent of the filaments, such a rotation implies the formation of a pair of $\pm 1/2$ disclinations at the point of detachment between the two filaments (Figure \ref{fig_SingleBuckling}H). This process critically relies on filament self-adhesion, here implemented as a short-ranged attractive potential between beads in a single chain. A purely repulsive, excluded-volume, interaction would limit the self-contact region of a filament to a single point. By contrast, addhesion extends this region to a finite fraction of the filament length, thus ``locking'' the buckling-induced rotation of the nematic director and preventing the newly unbound defects from an immediate annihilation. 
Furthermore, as shown in SI Sec. 2.3, at early stages of loop formation—when the loop area remains small and adhesive forces can be approximated as constant—a balance between adhesion and growth stabilises both loop size and shape. This transient stabilisation further protects $+1/2$  defects. At later times, however, secondary buckling events destabilise the loop, unlocking the structure and promoting defect annihilation.

\section*{Topologically patterned active turbulence}

After isolated bacterial chains elongate along the trap’s orientation, cells enter in contact with cells from neighbouring chains (Figure \ref{fig_Perp}A at $t=90$~min). Upon contact, local orientation is destabilized, starting a regime of active turbulence. In the following, we investigate, using both experiments and numerical simulations, two distinct trap configurations -- perpendicular and parallel -- and demonstrate how orientation not only controls the early-time dynamics of the active turbulent flow, but also the global nematic order in the steady state. Crucially, we show that tuning the initial trap separation can lead to collective filament buckling, which alters the previously described unbinding of $\pm 1/2$ topological defects from single-filament evolution. 

In experiments, we acquire time-lapse microscopy images of the growing bacterial films with different initial trap spacing and orientation. We analyse the time-lapses to measure the evolution of the local chain orientation, i.e. the nematic director field (Figure \ref{fig_Perp}A in green) and to detect the number, position and orientation of topological defects (shown in Figure \ref{fig_Perp}A as cyan arrows and red trefoils, Supplementary Movie 2-6). For experimental data representation, the time axes are offset (depending on trap density) so that all time series reported below correspond to the same bacterial density at the start of the observation period (see Methods). 
 
\subsection*{Perpendicular orientation does not affect the active nematic}\label{sec3}

\begin{figure}[h]
\centering
\makebox[\textwidth][c]{
    \includegraphics[width=\textwidth]{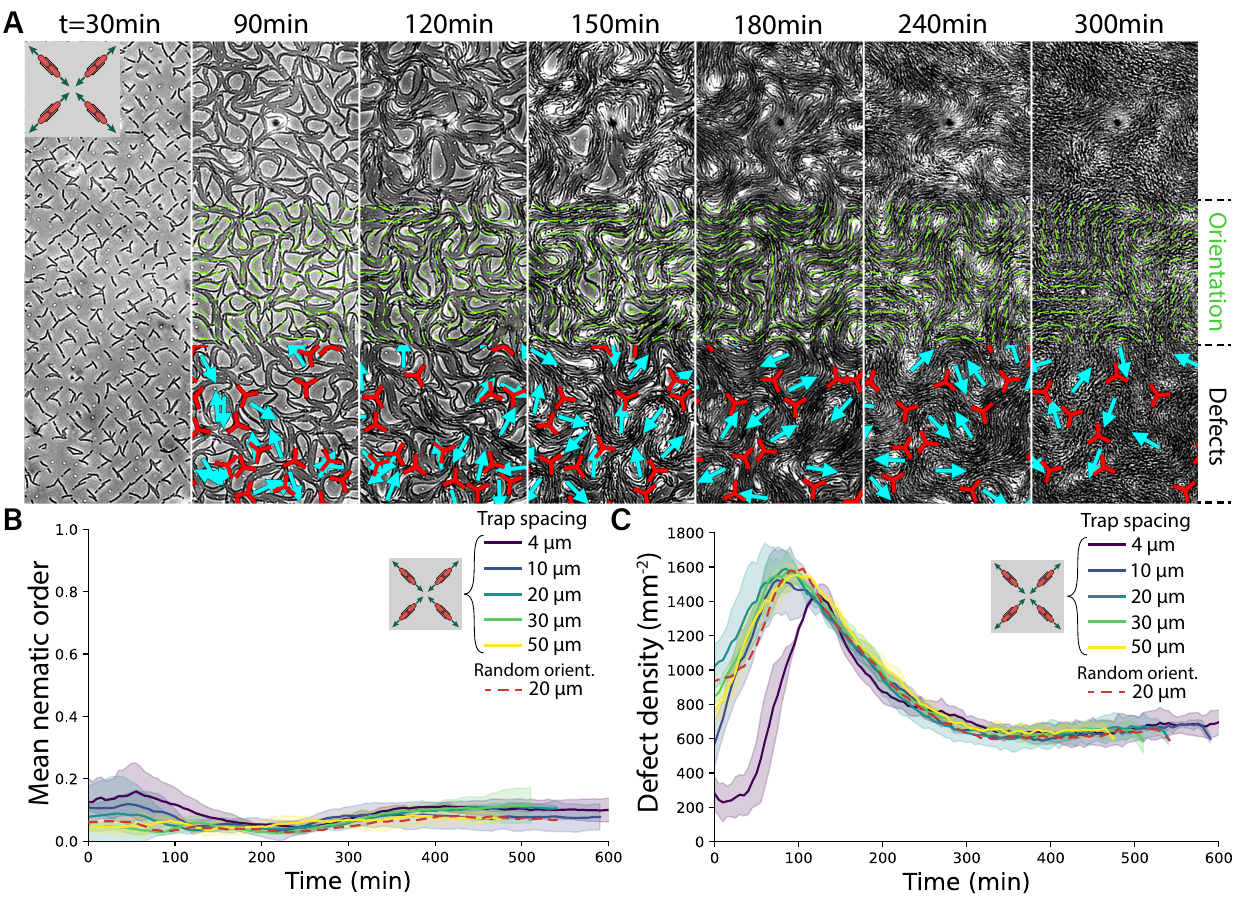}
    }
\caption{\textbf{Fig. 3 Nematic order and collective flows emerging from elongating \textit{B. subtilis} chains with perpendicular initial orientation.} (A) Time-lapse images of growing \textit{B. subtilis} bacterial chains starting from v
$10 \mu$m-spaced spore arrays with perpendicular outgrowth orientations at t~=~30, 90, 120, 150, 180, 240 and 300~min. The bright-field images are analysed to reconstruct the nematic director field, corresponding to local filament orientation, superimposed in green on the middle part of the images. The director field is used to detect +1/2 (cyan arrow) and -1/2 (red trefoil) topological defects. Time evolution of the mean nematic order (B) and defect density (C) for different trap spacing with perpendicular orientation. Lines represent the mean and shaded areas represent $\pm$one standard deviation for three biological replicates. The red dashed line represents the corresponding time evolution for randomly oriented traps with $20 \mu$m spacing.}\label{fig_Perp}
\end{figure}

Perpendicular trap configurations do not impose a lasting orientational order, and the system rapidly evolves into an active turbulent state characteristic of extensile active nematics. As a representative example, the time-lapse for perpendicularly-oriented traps with a $10 \mu$m spacing (Figure \ref{fig_Perp}A) displays active flows with the typical features of active turbulence 
~\cite{giomi_defect_2014}. The same amount of $+1/2$ and $-1/2$ defects is present, where only the positive defects are motile (Figure S3) and move in the direction of the cyan arrow (Figures \ref{fig_SingleBuckling}G,E). After 6-8~h, however the system slows down drastically as cellular crowding inhibits further proliferation, leading to strong reduction of both flows and defect velocities (Figure S3, S4).

Irrespective of the initial trap spacing between 4 to $100 \mu$m, for perpendicular traps, the nematic order remains low throughout the initial growth phase, while topological defects exhibit a proliferation burst in the first 100 min and eventually plateau at a finite density (Figure \ref{fig_Perp}B,C). Both the maximal and asymptotic defect density, as well as the dynamics of the process, are very similar across all trap spacings and match those of randomly-oriented traps spaced by $20 \mu$m. This indicates that, in perpendicular configurations, initial seeding conditions do not significantly influence the resulting active nematic state .

\subsection*{Parallel orientations force high order and induce synchronous buckling}\label{sec4}

\begin{figure}[H]
\centering
\makebox[\textwidth][c]{
    \includegraphics[width=\textwidth]{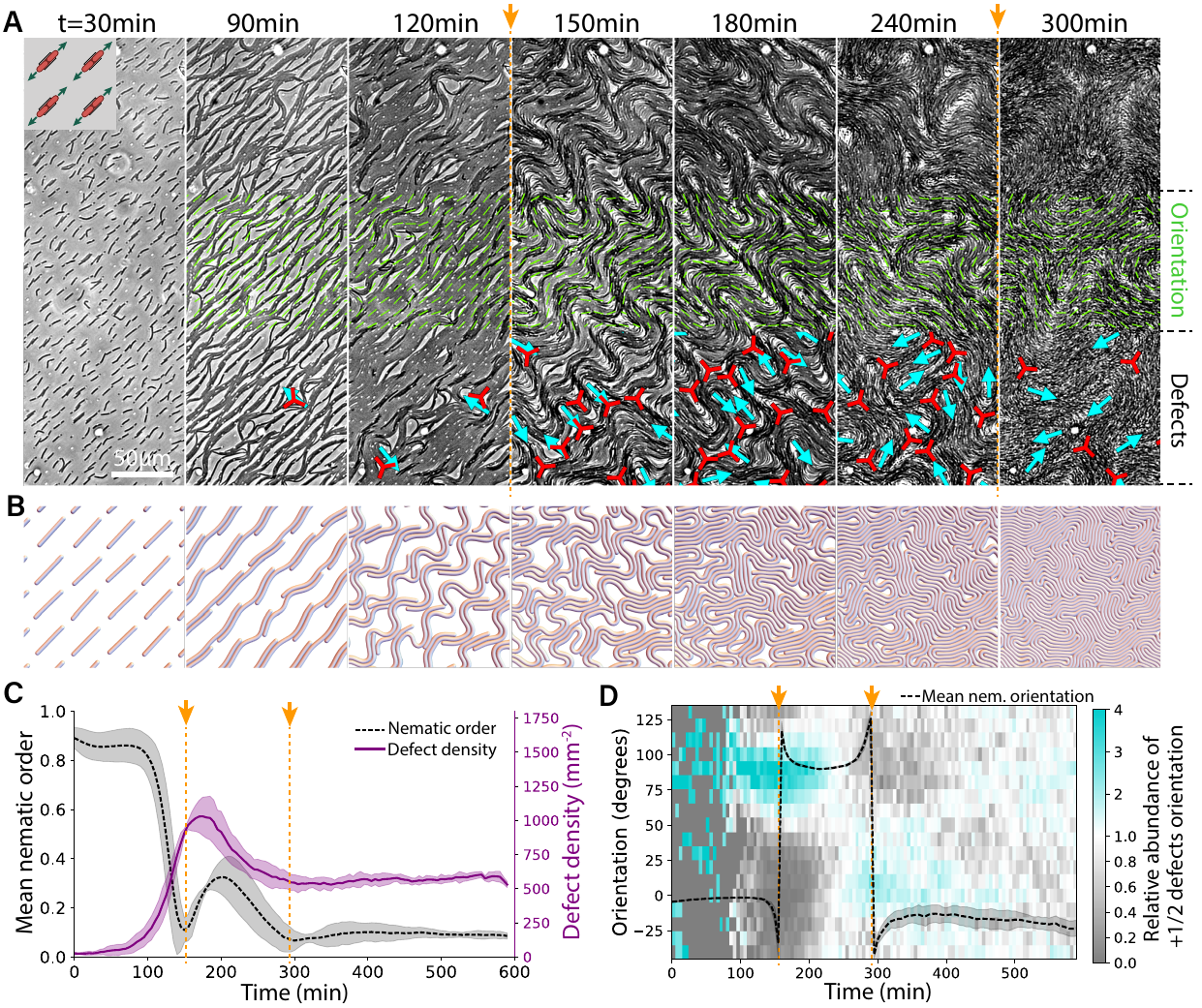}
    }
\caption{\textbf{Fig. 4 Parallel initial orientation influences \textit{B. subtilis} active nematic by imposing early high nematic order.} (A) Time-lapse images of \textit{B. subtilis} bacterial chains grown from $10 \mu$m spaced spore arrays with parallel orientation at t~=~30, 90, 120, 150, 180, 240 and 300~min. The nematic director field corresponding bacteria orientation is superimposed in green on the middle part of the images and is used to detect +1/2 (cyan arrow) and -1/2 (red trefoil) topological defects. Collective buckling events are represented by orange arrows and dashed lines, corresponding to the time points 150~min and 290~min. (B) Snapshots from a multi-chain numerical simulation at different times, from the initial growth until the emergence of the dense nematic phase. (C) Time evolution of the mean nematic order (dashed black line) and defect density (purple solid line) for $10 \mu$m spaced trap with parallel orientations. Lines represent the mean and shaded areas represent $\pm$one standard deviation for three biological replicates. (D) Time evolution of the mean nematic orientation (dashed black line) and +1/2 defect orientation (heat map for defect abundance relative to a uniform distribution) for $10 \mu$m spaced trap with parallel orientations. Grey corresponds to depletion and cyan corresponds to enrichment of defect orientation compared to the uniform distribution, represented coloured in white. Lines represent the mean and shaded areas represent $\pm$one standard deviation for two technical replicates. The orientation axis is wrapped to account for the $\pi$-symmetry of the nematic director; sharp vertical features in the mean orientation curve represent rapid but continuous director rotations rather than physical discontinuities. }\label{fig_Para}
\end{figure}

By contrast, patterning spores with a parallel orientation leads to a distinctively different behaviour. For parallel-oriented traps with a $10 \mu$m spacing, time-lapse images show an almost perfect nematic order at early times, where the unbinding of a small number of defects originates from the isolated buckling of single filaments across the system (Figure \ref{fig_Para}A,C , $t=120$~min). As the filaments keep growing, we then observe collective buckling where multiple pairs of $\pm 1/2$ simultaneously unbind across the system, as marked by a sharp drop in the nematic order parameter(Figure \ref{fig_Para}A, $t=150$~min). This collective buckling instability, is further characterised by the emergence of large-scale wavy patterns, known as {\em ripplocations}, which are also observed in systems of fibres constrained within a matrix or mechanically coupled to adjacent fibres~\cite{li_instabilities_2018, barsoum_ripplocations_2019}. Both phenomena are reproduced by our numerical simulations, which confirms that all the microscopic mechanisms that are instrumental to the emergence of collective buckling are built in our model of individual filaments (Figure \ref{fig_Para}B).

In both experiments and simulations, topological defects emerge at each buckling loop’s location as pairs of positive and negative defects oriented orthogonally ($\pm 90 ^\circ$) to the initial director field (Figure \ref{fig_Para}A-B, $t=150\text{--}180$~min). Growth further separates the +1/2 defects along this orthogonal direction, away from their corresponding $-1/2$ defect (Figure \ref{fig_Para}A, $t=180$~min). This concerted defect motion redirects the mechanical stresses and reorients the mean nematic director along the $+1/2$ defects’ orientation with a rebound of the nematic order to intermediate values of $\sim 0.3$ (Figure \ref{fig_Para}C, D). In these conditions, a second synchronous buckling event takes place, further releasing the compressive stresses in the orthogonal direction by reorienting the topological defects to $90^{\circ}$ or $-90^{\circ}$ again. However, this second event leads to a weaker defect alignment compared to the first buckling (Figure \ref{fig_Para}A,D , $t=240\text{--}300$~min). In comparison, perpendicular or randomly oriented traps lack preferential defect orientation and show no collective reorientation events (Figure S5,S6).

\section*{Nematic order is lost with larger trap spacing}\label{sec5}

\begin{figure}[H]
\centering
\makebox[\textwidth][c]{
    \includegraphics[width=\textwidth]{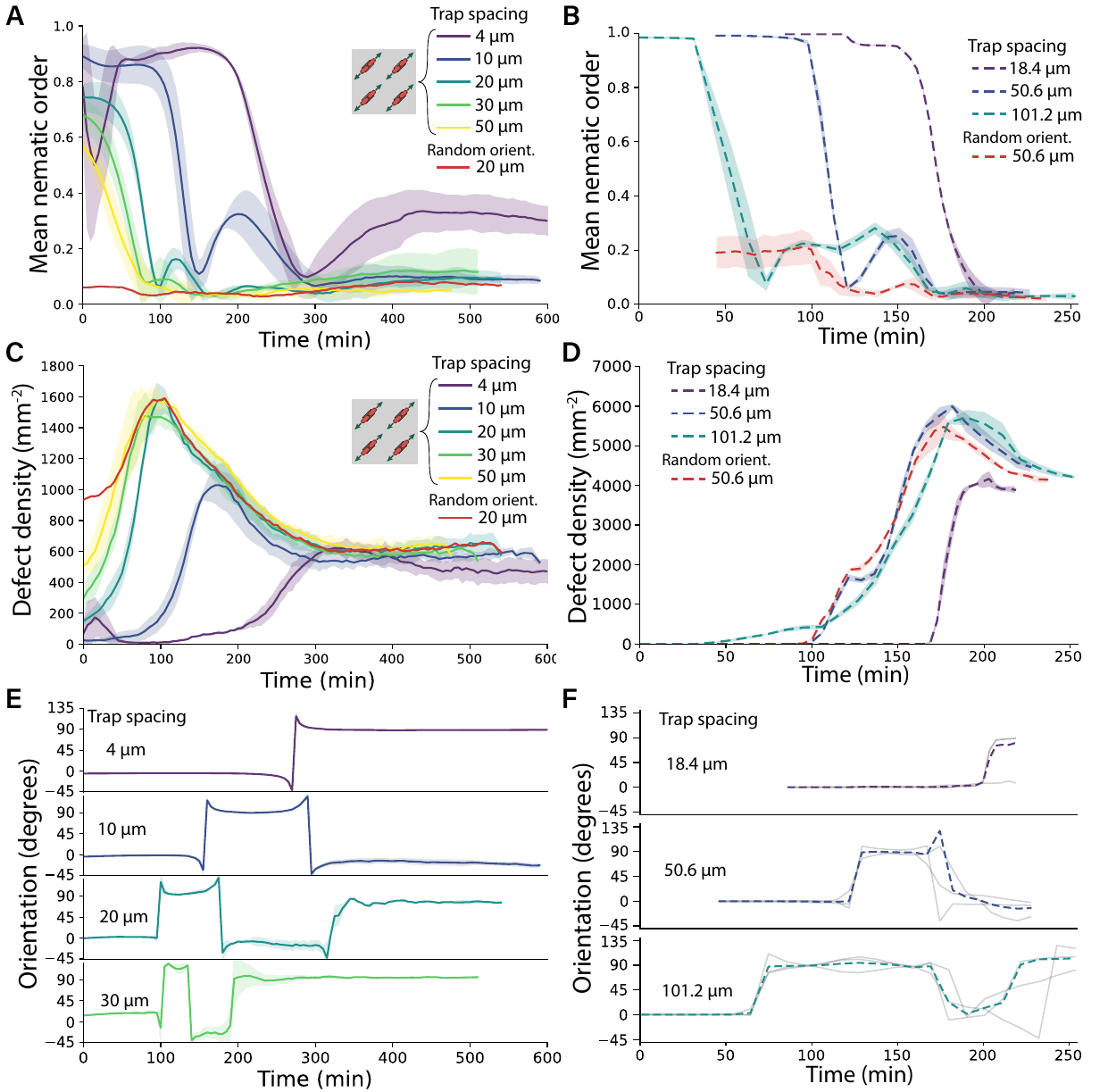}
    }
\caption{\textbf{Fig. 5 Influence of trap spacing on nematic order, defect density and their orientations for parallel initial configurations.} (A) Time evolution of the mean nematic order and (C) defect density for parallel orientations. Shaded areas represent $\pm$one standard deviation for three technical replicates. Red dashed line represents the corresponding time evolution for randomly oriented traps with 20$\mu$ spacing. (B, D) Corresponding simulation results for the mean nematic order and defect density. (E) Time evolution of the mean nematic orientation for 4, 10, 20 and 30 $\mu$m spaced trap with parallel orientations. The larger the spacing the more synchronous buckling events. Lines represent the mean and shaded areas represent $\pm$one standard deviation for two technical replicates. (F) Corresponding simulation results for the mean nematic orientation. Replicates are represented as grey lines. The orientation axis is wrapped to account for the $\pi$-symmetry of the nematic director; sharp vertical features in the mean orientation curve represent rapid but continuous director rotations rather than physical discontinuities.}\label{fig_MultiBuckling}
\end{figure}

Increasing trap spacing, relative to the $10 \mu$m case, while maintaining parallel orientation reduces the overall propensity to develop nematic order. Specifically, the highly ordered configurations observed at early times are now less persistent, and $\pm 1/2$ defects unbind at a rate which increases with the trap separation. Consequently, larger spacing induces behaviour that becomes more similar to the one observed for the perpendicular or random configurations (Figure \ref{fig_MultiBuckling}A, C). These findings demonstrate that the initial orientation of the traps only influences the active nematic phase if their spacing is sufficiently small relative to the critical self-buckling length, $L_c$ (Fig. \ref{fig_SingleBuckling}). Since filaments grow out at both ends of each trap, neighbouring filaments first come into contact at half the inter-trap distance . Given the trap positioning and orientation on a square lattice, this midpoint corresponds to half the length of the lattice diagonal. Consistently with the measured $L_c \approx 90 \mu$m  in our system, we observe that nematic order begins to deteriorate when traps are spaced more than $\sim 30 \mu$m (diagonal $\sim 45 \mu$m). In this case, filaments undergo spontaneous self-buckling before contact with other filaments and their stochastic buckling overrides the guidance provided by the geometry. 
Our simulations recapitulate the main experimental trends, showing a similar shift in the timing of nematic order loss and nematic order rebound (Figure \ref{fig_MultiBuckling}B), and a reduction in defect appearance for small trap spacing (Figure \ref{fig_MultiBuckling}D). As in other buckling phenomena, however, disorder -- e.g. resulting from the dispersion of $L_{c}$ (see Figure \ref{fig_SingleBuckling}D,E) -- anticipates the instability compared to the ideal case~\cite{budiansky_theory_1974,li_mechanics_2012} or the simulations, thereby triggering premature buckling events. At higher densities, effects of disorder are instead outweighed by mechanical interactions.

The timing and number of synchronous buckling events also depend on the trap spacing. We observe a single buckling event for $4 \mu$m spaced traps, two for $10 \mu$ and three for 20 and $30 \mu$m (Figure \ref{fig_MultiBuckling}E). The larger the trap spacing is, the shorter the time between consecutive bucking events and the weaker the overall defect orientation alignment. Furthermore, the mean nematic orientation plateau values and timing of buckling events exhibit larger variations across biological replicates, indicating that if cells have more space to grow, the degree of imposed control is gradually lost (see Figure S6). Above a spacing of $30 \mu$m, there is essentially no preferential orientation of the nematic order and the system behaves analogously to those with initial perpendicular or random configurations. The number of buckling events as a function of trap spacing obtained from the simulations agrees very well with the experimental ones (Figure \ref{fig_MultiBuckling}F).

\newpage

\section*{Nematic bacterial films as living diffraction gratings }\label{sec6}

The understanding of the microstructural evolution of our bacterial films sets the basis for their use as optical living materials. Starting from parallel traps with small spacing, as cells get closely packed into strongly nematic films, they form highly-ordered structures with a $\approx 1 \mu$m periodicity, corresponding to the width of the \textit{B. subtilis} body. Upon transmission illumination with white light, this regular nematic arrangement acts as a diffraction grating, generating characteristic rainbow-like and angle-dependent colouration (or iridescence). 
Iridescence has been observed in different bacteria, but often only at the edge of macroscopic colonies, while the central parts appear white because disordered cell orientations favour light diffusion over coherent diffraction\cite{kientz_iridescence_2012}. Our degree of control over the alignment of \textit{B. subtilis} cells allows us to precisely tailor optical anisotropy within biofilms with local high nematic order.

\begin{figure}[H]
\centering
\makebox[\textwidth][c]{
    \includegraphics[width=\textwidth]{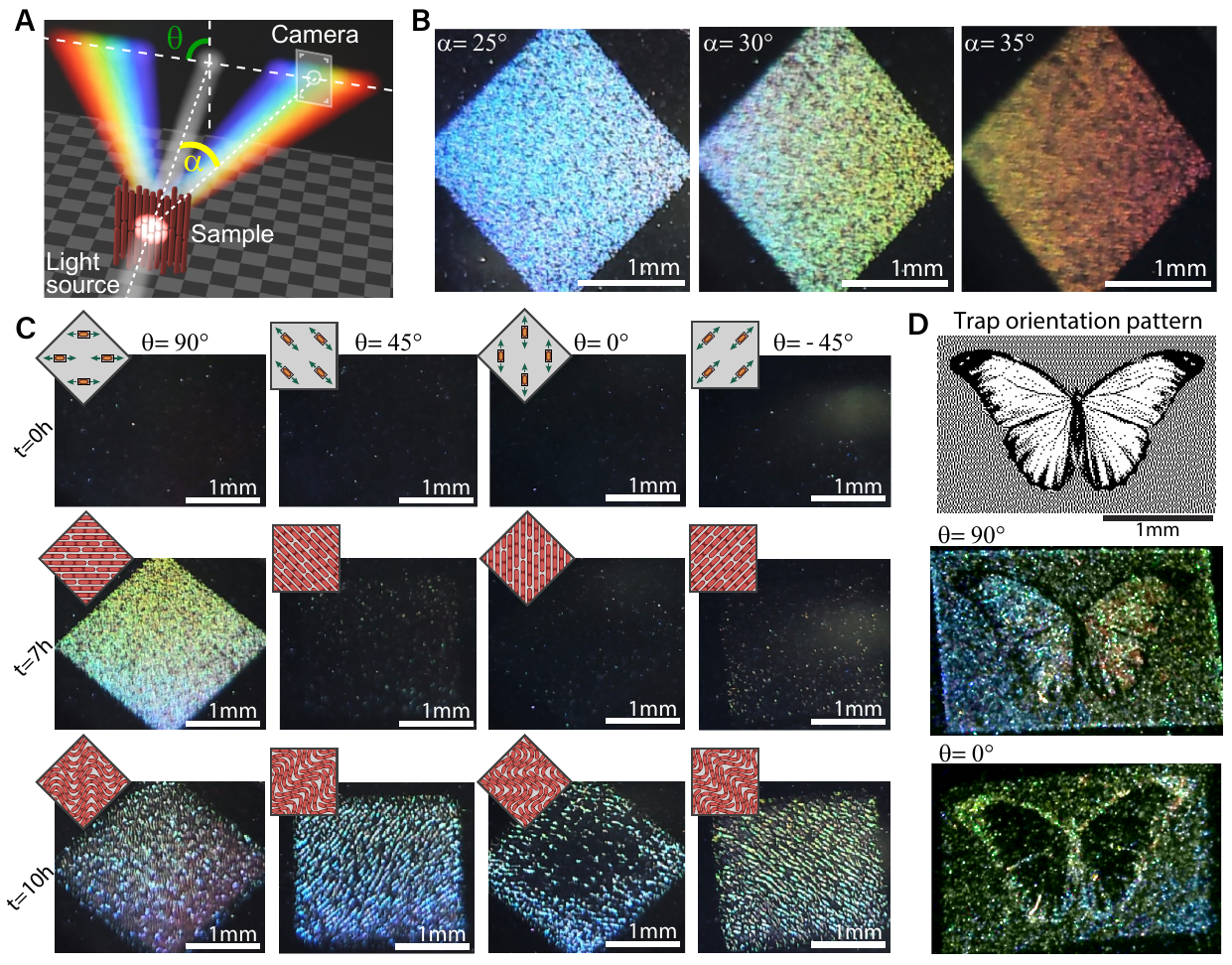}
    }
\caption{\textbf{Fig. 6 Ordered bacteria act as living diffraction gratings: controlling bacterial microstructure allows creating structural colouration patterns.} (A) Scheme of the observation geometry of the nematic films, showing the bacterial chains aligned vertically ("Sample"). White light with normal incidence to the sample is diffracted when passing through the bacterial film. Constructive and destructive interferences result in rays of light diverging along the normal plan to the bacteria orientation, with an angle $\alpha$, which depends on the light wavelength. The diffracted rays correspond to the structural colourations and are imaged with a camera. The camera can be tilted at different angles $\alpha$ and the sample can be rotated at different angles $\theta$ (between the trap orientation and the camera axis). (B) Images of the bacterial film at different angles $\alpha$, grown from a 4~$\mu$m spaced parallel trap array for $\sim$7~h of growth at RT. The larger the angle $\alpha$ the larger the wavelength of the structural colour.
(C) Images of nematic bacterial films at different angle $\theta$~=~90$^\circ$, 45$^\circ$, 0$^\circ$, -45$^\circ$ at t~=~0~h, 7~h and 10~h. At t~=~0~h, spore-loaded PDMS traps arrays do not exhibit any structural colour. At t~=~7~h, structural colour is only observed at $\theta$~=~90$^\circ$ as expected for a diffraction grating, corresponding to a quasi-perfect nematic order. At t~=~10~h, buckling of the chains occurred and structural colouration patterns are visible at all orientation angles $\theta$. Bright areas correspond to local orientation of chains at 90$^\circ$ with the camera axis while dark ones correspond to all other orientations. (D) Patterning of structural colours by controlling trap orientations. A binary image of a butterfly is used as template, where black pixels and white pixels correspond to 2$\times$2 arrays of traps with a SO-NE and a NO-SE orientation respectively (90$^\circ$ difference). When imaged at the quasi-perfect nematic order state ($\sim$7~h of growth), strong structural colouration emerges at $\theta$~=~90$^\circ$ for the positive and $\theta$~=~0$^\circ$ for the negative version of the image.}\label{fig_StructuralColor}
\end{figure}

We image a 2D film of growing bacteria seeded with 4~$\mu$m-spaced parallel traps using a multi-axis motorised microscope, which allows rotating the camera observation angle $\alpha$ relative to the sample (Figure \ref{fig_StructuralColor}A). We observe strong iridescent colouration from violet-blue at $\alpha$~=~25$^\circ$ to orange-red at $\alpha$~=~35$^\circ$(Figure \ref{fig_StructuralColor}B, $\theta$~=~90$^\circ$). The colours are caused by first-order diffraction, with increasing wavelengths diffracted at increasing angles $\alpha$. Such light diffraction does not originate from the microfabricated PDMS traps, as shown by the absence of any colour before film growth (Figure \ref{fig_StructuralColor}C, t~=~0h).
Based on the diffraction grating equation $d \sin{\alpha}=\lambda$, we estimate the grating spacing at d$\sim$1~$^\mu$m, which corresponds well to the cell chain spacing measured by microscopy (Figure S7). Importantly, such angle-dependent light diffraction can only be observed along the plane orthogonal to the nematic director field following bacteria orientation, i.e. with $\theta$~=~90$^\circ$ (Figure \ref{fig_StructuralColor}A). High nematic order bacterial films imaged at $\theta$~=~-45$^\circ$, 0$^\circ$ or 45$^\circ$ do not exhibit structural colouration (Figure \ref{fig_StructuralColor}C, t~=~7h, Supplementary Movie 7). Moreover, the previously described synchronous buckling leads to the formation of large scale elongated domains with local cell alignment along different directions (iridescence stripes for $\theta$~=~-45$^\circ$, 0$^\circ$, 45$^\circ$ -- Figure \ref{fig_StructuralColor}C, t~=~10h). 

We can harness our high-resolution local control of bacterial nematic order to create binary images in structural colouration. We microfabricate a 2D pattern of a butterfly composed of pixels containing 2$\times$2 traps, oriented 90$^\circ$ relative to each other for white and black pixel respectively. Upon bacteria growth along the imposed direction we then generate structural coloured pixels that are visible at either $\theta$~=~90$^\circ$ and 0$^\circ$, respectively (Figure \ref{fig_StructuralColor}D and Supplementary Movie 8). The local nematic order can also be observed by crossed polarised-light microscopy, with local extinction where chains are oriented orthogonally to the polariser (Figure S8). Once the cells grow to the point where high nematic order is lost, the image ceases to be visible. 

\newpage

In conclusion, the direct patterning of bacterial spores using capillary assembly offers exciting opportunities to seed, direct and control the growth of bacterial films with a high spatial resolution. By prescribing initial cell proximity and orientation, we can deterministically program local nematic order, offering an alternative to the spontaneous emergence of turbulent flows typical of active nematics~\cite{doostmohammadi_active_2018, yaman_emergence_2019, head_spontaneous_2024}.
Our results reveal that the macroscopic statistical properties of active turbulence, such as defect densities and their evolution, are robust across nearly all initial configurations --in particular any configuration where the spores are not parallel and not uniformly spaced below the buckling critical length-- and are connected to the single-filament mechanics and interactions, as confirmed by numerical simulations . 
Since active turbulence in bacterial suspensions has been shown to enhance both cell mixing~\cite{schwarzendahl_active_2022} and the transport of nutrients~\cite{sommer_bacteria-induced_2017}, it is tempting to speculate that similar turbulent-like dynamics in surface-attached bacterial films may also promote lineage mixing within growing colonies and potentially enhance the access to nutrients. These combined genetic and transport-related advantages suggest that entering this active state may be beneficial during colony growth. Our results indicate that single-filament mechanical properties may play a key role in leading to robust active turbulent states, despite variability in the initial cell distribution. This hypothesis could be tested by systematically varying filament mechanical properties, for example by using strains with different chaining behaviour or by tuning interactions with the surrounding medium~\cite{yaman_emergence_2019}. 

In contrast, the tightly-packed parallel configuration represents a state of broken rotational symmetry that is rarely realised in nature, bypassing active turbulence in favour of large-scale cellular alignment and spatially periodic, collective buckling patterns. Further exploration of similar initial configurations, therefore offers exciting opportunities to tailor local anisotropy toward the design of functional living materials with programmable mechanical, chemical, and optical properties. 

Finally, cell patterning with single-cell resolution on position and orientation offers tremendous possibilities to investigate the role of spatio-temporal organisation of bacterial colonies on microbial fate. Extending this methodology toward the deposition of multiple bacterial strains or species into prescribed spatial arrangements would allow for the systematic study of spatial  interactions\cite{boggon_single-cell_2023,henderson_disentangling_2025, warrier_interplay_2026,backer_spatial_2026}, such as competition and cooperation\cite{le_bec_optogenetic_2024}, with unprecedented resolution. While the use of bacterial spores restricts the choice of the working organisms, the robustness of the method and the possibility of genetic engineering of the vegetative cells growing from the spores open up a vast range of future studies.

\section*{Methods}\label{sec7}

\subsection*{Bacterial strains and sporulation protocols}\label{subsec7.1}
Experiments were conducted using the \textit{Bacillus subtilis} strain 168~\cite{setlow_role_1996}, generously provided by Dr. Rosa Heydenreich and Prof. Dr. Alexander Mathys. This strain is non-motile and only produces cell chains when growing on solid substrates.
Spores were prepared using Difco sporulation medium agar (pH 7.6) following the protocol of the protocol of Zhang et al. (2020)\cite{zhang_flow_2020}. After 4 days at 37~$^\circ$C, spores were harvested, washed 4$\times$ with MilliQ water by centrifugation (6000~g, 10~min, 4~$^\circ$C). Spore’s purity (absence of vegetative cells) was checked under Phase Contrast microscopy. Batches of spores with low purity (under ~90~\%) were further purified by density centrifugation: spore suspensions were slowly pipetted on top of a 50~\% w/v Nicodenz solution and centrifuged at 6000 g for 10~min. The upper liquid was discarded and the spores in the pellet were recovered and washed twice with MilliQ. Spores were washed twice per day in the following 7 days after harvest. Spores were stored at 4~$^\circ$C in MilliQ water and were washed every two weeks.

\subsection*{Microfabrication of the trap arrays in PDMS}\label{subsec7.2}

The PDMS templates for spore deposition were produced by replica moulding. The master moulds were fabricated by two-photon lithography on 2.5$\times$2.5~cm fused silica substrates using a Nanoscribe Photonic Professional GT2 (Nanoscribe GmbH). The photoresist IP-Dip and a 63$\times$NA 1.4 objective were used with default printing parameters. The printed master moulds were immersed into PGMEA for 20~min and rinsed with isopropyl alcohol to wash the unpolymerised photoresist. The master moulds were then placed in a 365~nm UV oven for 8~h to enhance the adhesion between polymer and substrate. The master moulds were fluorosilanised by chemical vapour deposition of fluoro-octyl-trichloro-silane for 1~h under vacuum and subsequently rinsed with ethanol before replica molding.

Each printed cuboid trap is 2.0$\times$1.1$\times$0.8~$\mu$m (l$\times$w$\times$h) and they are arranged into arrays of 2$\times$2~mm. The array follows a square lattice with the traps oriented at 45$^\circ$ or 135$^\circ$ relative to the lattice orientation to obtain arrays in a parallel(only 45$^\circ$)  or perpendicular configuration (alternating 45$^\circ$ and 135$^\circ$). We used trap spacing of 4, 10, 20, 30 and 50~$\mu$m for the arrays, and of 100~$\mu$m for the isolated cell chain buckling experiments.

PDMS templates were fabricated using a PDMS precursor and curing agent in 10:1 weight ratio (Sylgard 184). Liquid PDMS was poured onto the master moulds and degassed for 30~min. The PDMS templates were polymerised overnight in an oven at 70~$^\circ$C and were detached from the master moulds to carry out spore deposition.

PDMS roofs for depositions were obtained in the same manner with 3D printed moulds (resin lithography Prusa SL1).

\subsection*{Spore deposition and growth conditions}\label{subsec7.3}

The PDMS assembly used for spore deposition is built by attaching a PDMS template with the trap array to a PDMS roof with a 0.12~mm thick SecureSeal™ double sided tape and placed onto a glass slide (see Figure S1 for the final assembly). The spore suspensions for deposition were prepared as follows: 200~$\mu$L of stock spore solution was washed with 50~mM Tris buffer (2~min at 16 000 g) and resuspended in 200~$\mu$L of 0.1~\% v/v Twin-20 50~mM Tris buffer. The Tris buffer was not pH balanced to favour high pH, which prevents spore aggregation. The spore suspension was diluted with 0.1~\% v/v Twin-20 50mM Tris buffer to reach a spore concentration of ~1E9 CFU/mL. 30~$\mu$L of the spore suspension was injected in the PDMS assembly through the loading port using a pipette. The PDMS assembly was then placed on a heat plate at 30 $^\circ$C and left for around 2 hours to allow the droplet to evaporate and deposit the spores in the traps. After the deposition was complete and the droplet dried, the PDMS assembly was disassembled to recover the PDMS template with spore-loaded trap arrays. After the excess PDMS was cut out, the loaded template was placed into a Nunc™ glass bottom dish (4~cm of diameter). Tryptone soy broth (TSB) with 1~\% w/v gellan gum (Gelzan™) was melted in a microwave and allowed to equilibrate in a 75$^\circ$C water bath. Using a serological pipette, 3~mL of the melted TSB-gellan gum was poured onto the spore-loaded PDMS template and allowed to cure for 5~min lid open inside a biosafety cabinet. The dish was closed with the lid to prevent drying while keeping a small gap for gas exchange and was then transferred into the microscopy setup. The stage and focus positions were configured and the timelapses were started about 15~minutes after gel curing.

\subsection*{Microscopy imaging}\label{subsec7.4}

Bright field images were captured every 5~minutes using an inverted Nikon Eclipse Ti2-E epifluorescence microscope, equipped with a Nikon objective 20x/0.45 S Plan Fluor ELWD, a Orca-flash4.0 Hamamatsu camera, a Perfect focus system (PFS) and a controlled temperature box set at 30$^\circ$C. 
The 2$\times$2 mm trap arrays were imaged only over a central area of 1.2$\times$1.2 mm to avoid edge effects (4 different positions with a field of view of 0.6$\times$0.6 mm). Technical replicates correspond to two positions within the same trap array. Biological replicates correspond to different trap arrays deposited separately and imaged separately.

\subsection*{Structural colour imaging}\label{subsec7.5}

The structural colouration of bacterial films was captured with a Microqubic MRCL700 3D imager, which offers multiple degrees of freedom for sample and camera motion. The device was set-up as a transmission microscope, illuminating through the sample from below with white light and imaging from above. The camera was rotated out of the vertical axis while maintaining the sample within the field of view and in focus. This rotation allows to observe the sample at different angles $\alpha$ to image the different wavelengths of the diffracted light. The sample was also rotated around its centre using a 3D printed rotating holder at different angles $\theta$ (relative to the trap orientation) to investigate the directionality of the bacterial diffraction grating. The 3D imager was programmed to capture multi-orientation time-lapses autonomously using a Python script. Note that different colours are simultaneously observed on each image because the $\alpha$ value varies between the optical paths of the left and the right side of the camera. These experiments were run at room temperature, which slows bacteria growth without affecting the overall behaviour.

\subsection*{Image analysis}\label{subsec7.6}
Topological defects are detected using the following image analysis workflow on time-lapses of 12~h of bacterial growth, imaged every 5~min with bright field microscopy.

For each image, the first step is finding the orientation field using ImageJ plugin “OrientationJ”\cite{puspoki_transforms_2016} (\url{https://bigwww.epfl.ch/demo/orientationj/}) with the parameters: local window sigma~=~5 pixels, Gradient~=~Cubic Spline, Grid size~=~4 pixels. The nematic order field is then computed and +1/2 and -1/2 topological defects location and orientation is determined using a topological defect detection algorithm based on machine learning\cite{killeen_machine_2024} (\url{https://github.com/KilleenA/ML_DefectDetection}). When the entire stack is processed, the +1/2 and -1/2 defects are tracked independently using TrackPy with the parameters: maxdistance~=~40, minimum tracked frames~=~3, skip memory~=~1. The defects that are not tracked, e.g. that are not detected for three consecutive frames, are considered as noise and are removed from the analysis. The tracked topological defects are used to retrieve the variation of defect density, velocity and orientations with time.

The mean nematic order and the mean nematic orientation were computed using the orientation fields, weighted using the OrientationJ coherency coefficient which indicates whether the local 4$\times$4 pixel window features are oriented or not: 1 if the local structure has one dominant orientation and 0 if the image is essentially isotropic.

\subsection*{Data temporal alignment}\label{subsec7.7}
To account for variations in initial bacterial density arising from different trap spacings, experimental trajectories were temporally aligned by assuming an exponential growth model, $d(t) = d_{t0} e^{\lambda t}$, where $\lambda \approx 1.65 \text{ h}^{-1}$ represents the bacterial growth rate measured in Figure S2. To accurately synchronise the trajectories to an equivalent developmental stage across all conditions, we introduced a spacing-dependent temporal offset $\Delta t(s)$. This temporal alignment follows the logarithmic scaling:$$\Delta t(s) = \frac{2}{\lambda} \ln(s)$$
With $s$ the trap spacing length.

\subsection*{Numerical simulations}\label{subsec7.8}
Each chain is modelled as an over-damped sequence of $N$ beads connected by stiff springs, similarly to \cite{isele2015self,kurjahn2024collective}. The equation of motion for $i$-th particle can be summarised as:
\begin{equation}
\boldsymbol{\nu}_i \cdot \Dot{\boldsymbol{r}_i} = -\nabla_i U^{\left(el\right)} - \nabla_i U^{\left(bend\right)} - \nabla_i U^{\left(int\right)} - \boldsymbol{F}_i^{\left(fric\right)}, 
\end{equation}
where $\boldsymbol{\nu}_i$ is the anisotropic viscous drag tensor, $U^{\left(el\right)}$, $U^{\left(bend\right)}$ and $U^{\left(int\right)}$ the potential energies associated, respectively, to elastic compression/stretching, to bending stiffness and to pairwise interaction with other beads (belonging to the same or to different chains), taking in account both exclude-volume and adhesive effects, and $\boldsymbol{F}_i^{\left(fric\right)}$ is the effective friction force. The precise functional form of these terms is given in the Supplementary Information. Since the bacteria are non-motile we do not include an active self-propulsion force. We also do not consider thermal-like fluctuations, with all the noise coming from the initial condition. The system is driven out-of-equilibrium by growth: the natural length $l_i$ in the elastic potential $U_e$ increases with time according to a generalised logistic law.

\section*{Data availability}

All the data supporting the findings of this study are included in the article and its Supplementary Information file.  Additional data may be requested from the authors.

\section*{Code availability}

The code to run the simulation is available from the corresponding author upon request

\section*{Acknowledgements}

This work is partially supported by the ERC-CoG grant HexaTissue (G.P.M., L.P., L.G.).
Numerical simulations were performed using the ALICE compute resources provided by Leiden University.

\section*{Author information}

\subsection*{Authors and Affiliations}
\noindent\textbf{Materials Department, ETH Zürich, Zürich,Switzerland.}

Matthias Le Bec, Cameron Boggon, Yiyao Hu, Lucio Isa.

\noindent\textbf{Instituut-Lorentz, Universiteit Leiden, Leiden,The Netherlands.}

Guillem Pérez Martín, Leonardo Puggioni, Luca Giomi.

\noindent\textbf{Civil, environmental and Geomatic Engineering Department, ETH Zürich, Zürich, Switzerland.}

Eleonora Secchi.

\noindent\textbf{Health Sciences and Technology Department, ETH Zürich, Zürich, Switzerland.}

Rosa Heydenreich, Alexander Mathys.

\subsection*{Contributions}
Conceptualisation: M.L.B., G.P.M, C.B., L.P, L.G., E.S., L.I.; Data curation and Formal analysis: M.L.B., G.P.M, L.P.; Funding acquisition: L.G., L.I.; Investigation: M.L.B., G.P.M., L.P.; Methodology: Y.H., R.H.; Project administration: L.G., E.S., L.I.; Resources: R.H., A.M.; Software: M.L.B., G.P.M; Visualisation: M.L.B., G.P.M; Writing - original draft: M.L.B., G.P.M., L.P., L.G., E.S., L.I. ; Writing - review $\And$ editing: all authors.
\section*{Ethics declarations}
The authors declare no competing interests.

\section*{Peer review}

\section*{Additional information}

\section*{Supplementary information}

Supplementary information

Supplementary Movie 1 (AVI, 21.1 MB): Isolated chain buckling

Supplementary Movie 2 (AVI, 316 MB): Bacterial chains grown from 4~$\mu$m spaced spore arrays with perpendicular or parallel orientations with topological defects superimposed.

Supplementary Movie 3 (AVI, 268 MB): Bacterial chains grown from 10~$\mu$m spaced spore arrays with perpendicular or parallel orientations with topological defects superimposed.

Supplementary Movie 4 (AVI, 255 MB): Bacterial chains grown from 20~$\mu$m spaced spore arrays with perpendicular or parallel orientations with topological defects superimposed.

Supplementary Movie 5 (AVI, 253 MB): Bacterial chains grown from 30~$\mu$m spaced spore arrays with perpendicular or parallel orientations with topological defects superimposed.

Supplementary Movie 6 (AVI, 233 MB): Bacterial chains grown from 50~$\mu$m spaced spore arrays with perpendicular or parallel orientations with topological defects superimposed.

Supplementary Movie 7 (AVI, 2.60 MB): Living diffraction grating growing from 4~$\mu$m spaced spore arrays with parallel orientations.

Supplementary Movie 8 (AVI, 1.55 MB): Binary living diffraction grating growing from a trap pattern of a butterfly image.

\section*{Rights and permissions}
\newpage

\begin{center}
    \vspace*{1cm}
    {\huge \textbf{Supplementary Information}} \\[0.5cm]
    {\Large Shaping nematic order in bacterial films with single-cell resolution patterning} \\[1cm]
\end{center}

\vspace{1cm}

\tableofcontents
\newpage

\section{Supplementary methods}\label{supsec1}

\subsection{Spore germination time and bacteria growth rate}\label{subsec1.1}

To segment single bacteria filaments, brightfield images are inverted, a background subtraction algorithm is used (radius 5 pixels) and an autothreshold is applied. The obtained segmented regions’ centroids are tracked using TrackPy (maxdistance=20, minimum tracked frames=5, skip memory=1). The evolution of the tracked regions’ areas is used to filter out erroneous segmented regions by only keeping regions with area growing exponentially (linear regressions of the natural logarithm of the area with coefficient of determination $R^{2}\geq 0.99$). The remaining bacterial area growth curves
$A(t-t_{0} )=A_{0} e^{(\mu(t-t_{0}))}$ are used to determine the growth rate $\mu$ and the germination time  $t_{0}=\frac{-ln(A_{0})}{\mu}$

\subsection{Single filament buckling critical length}\label{subsec1.2}
Single bacteria filaments are segmented as described previously (linear regressions of the natural logarithm of the area with coefficient of determination $R^{2}\geq 0.98$). We then use the FIJI plugin “ContourCurvature” (Bureau H. Glünder 2020-2022) with smoothing (smoothing parameter=6) to measure the curvature of the filament mask contours. The buckling timing is defined as the time when the negative curvature of the tracked filament exceeds a value of 0.2. We then use the FIJI plugin skeletonize\cite{polder_measuring_2010} to measure the filament length at the buckling time, which is also used to filter out contour errors due to segmentation by verifying the skeletonized filament is composed of a single branch. We finally filter out critical length values below 15~µm which corresponds to filaments already buckled at very early stage that are not representative of the buckling phenomenon.

\subsection{Particle Image Velocimetry image analysis}\label{subsec1.3}
Liquid-like flows of cells powered by bacterial growth are quantified using Particle Image Velocimetry (PIVlab matlab plugin\cite{thielicke_pivlab_2014}) to retrieve flow field components such as velocity, vorticity and strain rate. The PIVlab parameters are the following: [Pre-processing:Enable CLAHE, window size 64 pix, Auto contrast stretch ; PIV settings: Algorithm multipass FFT window deformation, First pass interrogation area 64 pix step 32 pix, 2nd pass interrogation area 32pix step 16pix, Sub-pixel estimator Gauss 2x3-point, Correlation robustness standard ; Vector validation: Standard deviation filter Threshold 8, Local median filter Threshold 3, Interpolate missing data].

\subsection{Cell thickness and cell-cell distance measurements}\label{subsec1.4}
Cell thickness was measured using the intensity profiles of brightfield images of bacteria cluster of 1, 2, 4 and 6 parallel bacteria filaments. The total sizes of the cell clusters were measured between the two inflexion point of the intensity profiles. The cell thickness is estimated using the slope of the linear regression between the inflexion point distances and number of cells. The cell-cell distance was evaluated using an bright field image of a monolayer of bacteria filaments mostly parallel to each other (mean nematic order $>0.8$). A Fast Fourier Transform operation was performed on the image to detect the length scale of image feature. In particular, we compared the FFT radial profiles along two orientations: the profile 1 orthogonal to the cell orientation and profile 2 parallel to the cell orientation. The peak observed in profile 1 and in the normalised signal corresponds to the magnitude of the cell-cell distance.

\newpage
\section{Bending and looping in {\em B. subtilis}}

The buckling instability of an individual filamentary bacterium, such a {\em B. subtilis}, has been numerically investigated in Yaman {\em et al}.~ \cite{yaman_emergence_2019}. In the following, we provide an analytical solution of the buckling problem and  a detailed analysis of the post-transitional scenario, whose role is instrumental to the unbinding of $\pm 1/2$ nematic defect pairs.

\subsection{\label{sec:intro}Mechanical equilibrium of rods}

The simplest and more traditional approach to the elasticity of slender bodies, such as filaments or rods, is built upon the assumption that -- by virtue of the large separation between the scale of the longitudinal and transverse size of the body -- an arbitrary configuration of the latter can be captured by the sole geometry of its centerline: i.e. $\bm{R}=\bm{R}(s)$, where $0\le s \le L$, is the arc-length and $L$ the total length. In two dimensions, $\bm{R}$ traces therefore a plane curve, whose tangent and normal vector -- $\bm{T}$ and $\bm{N}$ respectively -- are given by 
\begin{equation}\label{eq:frenet}
\bm{R}'=\bm{T}\;,\qquad
\bm{T}'=\kappa\bm{N}\;,\qquad
\bm{N}'=-\kappa\bm{T}\;,
\end{equation}
where the prime indicates differentiation with respect to $s$ and $\kappa=\kappa(s)$ is the {\em signed} curvature of the curve. Eqs.~\eqref{eq:frenet} describe how the orthonormal frame $\{\bm{T},\bm{N}\}$ rotates along the curve. Once $\kappa$ is assigned in the form of a differentiable function, the fundamental theorem of plane curves guarantees that $\bm{R}$ is uniquely determined up to rigid motion. Furthermore, as in two dimensions the tangent vector depends on a single {\em turning} angle -- i.e. $\bm{T}=\cos\theta\,\bm{e}_{x}+\sin\theta\,\bm{e}_{y}$, with $\theta=\theta(s)$ -- using Eqs.~\eqref{eq:frenet} one has $\kappa=\theta'$ and 
\begin{equation}
\bm{R}(s) = \bm{R}(0)+\int_{0}^{s}\,\D{}s'\,\left[\cos\theta(s')\,\bm{e}_{x}+\sin\theta(s')\,\bm{e}_{y}\right]\;,
\end{equation}	
where $\theta(s)=\theta(0)+\int_{0}^{s}\D{}s\,\kappa(s')$. This construction implies $\kappa>0$ when $\bm{T}$ rotates counter-clockwise and $\kappa<0$ in the case of clockwise rotation.

Mechanical equilibrium can be described in terms of two vector fields expressing the resultant of the stresses and bending moments at play along a cross-section of the rod: i.e. $\bm{F}=\int \D{}A\,\bm{\sigma}\cdot\bm{e}_{\parallel}$ and $\bm{M}=\int \D{}A\,\bm{r}\times(\bm{\sigma}\cdot\bm{e}_{\parallel})$, where the integrals span the cross section of the rod, with $\bm{e}_{\parallel}$ the outward directed unit normal and $\bm{r}$ a planar position vector originating at the centre line. In these variables, the balance of forces and bending moments can be expressed via the equations
\begin{subequations}\label{eq:equilibrium}
\begin{gather}
\bm{F}'+\bm{f} = \bm{0}\;,\\
\bm{M}'+\bm{T}\times\bm{F}+\bm{m} = \bm{0}\;,	
\end{gather}
\end{subequations}
where $\bm{f}$ and $\bm{m}$ denote the external force and torque, respectively. Eqs.~\eqref{eq:equilibrium} must be complemented with a closure condition relating either $\bm{F}$ or $\bm{M}$ with the geometry of the centreline. For simplicity, in the following we will assume the classic Euler beam model (also known as Euler's {\em Elastica}), whose constitutive equation is given by
\begin{equation}\label{eq:moment}
\bm{M} = B\kappa\bm{e}_{z}\;,	
\end{equation}
with $B=EI$ the bending stiffness of the rod, $E$ the Young modulus and $I$ the areal moment of inertia. The choice of the sign in Eq.~\eqref{eq:moment} follows the same convention used for the curvature, thus a positive curvature, corresponding to a counter-clockwise rotation of the tangent vector requires a positive (i.e. parallel to the $z-$axis) bending moment and vice versa.

Eqs.~\eqref{eq:equilibrium} can be reduced to three scalar equations for the two components of the stress vector $\bm{F}$ and the curvature $\kappa$. Assuming $\bm{m}=\bm{0}$, taking $\bm{F}=F_{\parallel}\bm{T}+F_{\perp}\bm{N}$, $\bm{f}=f_{\parallel}\bm{T}+f_{\perp}\bm{N}$ and using Eqs.~\eqref{eq:frenet}, allows casting Eqs.~\eqref{eq:equilibrium} in the form
\begin{equation}\label{eq:scalar}
F_{\parallel}'-\kappa F_{\perp} + f_{\parallel} = 0\;,\qquad
F_{\perp}'+\kappa F_{\parallel} + f_{\perp} = 0\;,\qquad
B\kappa'+F_{\perp} = 0\;.
\end{equation}
From the left- and right-most equations, one can now calculate the stress resultant in the generic form
\begin{equation}\label{eq:stress_resultant}
\bm{F} = \left(T-\tfrac{1}{2}B\kappa^{2}\right)\bm{T}-B\kappa'\bm{N}\;,
\end{equation}
where 
\begin{equation}\label{eq:tension}
T(s)=\lambda-\int_{0}^{s}\D{}s'\,f_{\parallel}(s')\;,
\end{equation}
with $\lambda$ a constant, which is the tension across the rod. The central equation in Eqs.~\eqref{eq:scalar} yields instead an equation for the curvature $\kappa$: i.e.
\begin{equation}\label{eq:elastica}
B\left(\kappa''+\tfrac{1}{2}\kappa^{3}\right)-T\kappa - f_{\perp} = 0\;.
\end{equation}	
For $\bm{f}=\bm{0}$, $T=\lambda$ Eq.~\eqref{eq:elastica} reduces to the classic {\em Elastica equation} obtained from the minimisation of the energy functional 
\begin{equation}
E = \int_{0}^{L} \D{}s\,\left(\tfrac{1}{2}B\kappa^{2}+\lambda\right)\;.
\end{equation}
In this case, however, the tension $\lambda$ reduces to a Lagrange multiplier aimed at enforcing the incompressibility constraint $|\bm{R}'|=1$. Finally, in the special case where both $T$ and $f_{\perp}$ are uniform across the rod, integrating Eqs.~\eqref{eq:equilibrium} directly yields the following remarkably compact expression for the stress resulting: i.e. 
\begin{equation}\label{eq:F_vs_R}
\bm{F} = -f_{\perp}\bm{e}_{z}\times(\bm{R}-\bm{R}_{0})\;,
\end{equation}
with $\bm{R}_{0}$ an integration constant that one can adjust by suitably positioning the origin of the $xy$-plane with respect to the points along the rod where the stress resultant vanishes~\cite{capovilla_elastica_2002}. As long as $f_{\perp}\ne 0$, therefore, Eq.~\eqref{eq:F_vs_R} implies
\begin{equation}\label{eq:R_vs_F}
\bm{R}-\bm{R}_{0} 
= \frac{\bm{F}\times\bm{e}_{z}}{f_{\perp}}
= \frac{B\kappa'}{f_{\perp}}\,\bm{T}+\frac{T-\tfrac{1}{2}B\kappa^{2}}{f_{\perp}}\,\bm{N}\;,
\end{equation}
thus providing a general expression for the position of the rod, its curvature and the orientation of the local $\{\bm{T},\bm{N}\}$ frame.  

\subsection{\label{sec:buckling}Buckling instability}

To analyse how a buckling instability arises from drag, it is convenient to parametrise the centre line in Cartesian coordinates, so that $\bm{R}=x\bm{e}_{x}+y\bm{e}_{y}$, with $y=y(x)$ and $x=0$ marking the centre of the filament. This implies $\D{}s^{2}=\D{}x^{2}(1+y_{x}^{2})$, where we used the notation $y_{x}=\D{}y/\D{}x$ to highlight the difference with respect to the arc-length derivative $y'=\D{}y/\D{}s$. In these coordinates, one can readily show that
\begin{equation}\label{eq:cartesian}
\bm{T} = \frac{\bm{e}_{x}+y_{x}\,\bm{e}_{y}}{\sqrt{1+y_{x}^{2}}}\;,\qquad
\bm{N} = \frac{-y_{x}\bm{e}_{x}+\bm{e}_{y}}{\sqrt{1+y_{x}^{2}}}\;,\qquad
\kappa = \frac{y_{xx}}{(1+y_{x}^{2})^{3/2}}\;.
\end{equation}
At the onset of buckling, $|y_{x}|\ll 1$ and Eq.~\eqref{eq:elastica} can be approximated at the linear order in $y_{x}$. This gives
\begin{equation}\label{eq:small_grad}
By_{xxxx}-Ty_{xx} - f_{\perp} = 0\;.
\end{equation}
The classic Euler buckling occurs, in the absence of body forces, if the rod is subject to an axial compression so that $f_{\perp}=0$ and $T=-P$, where $P$ is the pressure applied at the tips. Eq.~\eqref{eq:small_grad} becomes then
\begin{equation}\label{eq:euler}
By_{xxxx}+Py_{xx} = 0\;.	
\end{equation}	
Assuming both ends free, so that $y(\pm L/2)=0$ and $y_{xx}(\pm L/2)=0$, using standard manipulations, one can show that the straight solution $y=0$ is unstable for $P/B>(\pi/L)^{2}$.

Now, to model deformations in {\em B. subtilis} we assume that each filament grows at constant rate and symmetrically from the centre of mass of the filament. The substrate, in turn, reacts to the elongation by means of a drag force
\begin{equation}\label{eq:drag}
\bm{f}_{\rm drag} =  -f\sign(x)\bm{e}_{x}\;,
\end{equation}
where $f$ is a constant proportional to the growth rate and the drag coefficient of the substrate. Thus, an arbitrary volume element located in the left half of the filament (i.e. $x<0$) experiences a positive drag force $\bm{f}_{\rm drag}=f\bm{e}_{x}$, while in the right half (i.e. $x>0$) the direction of drag is reversed and $\bm{f}_{\rm drag}=-f\bm{e}_{x}$. As we will show in Sec.~\ref{sec:looping}, self-adhesion between different portions of the same filament is instrumental to the formation of loops and the nucleation of $\pm 1/2$ defect pairs. However, because of the short-ranged nature of the adhesive interaction, this effect does not influence the shape of the filaments unless $|\kappa| \gg 1/L$ and the Euclidean distance between different portions of the same filaments becomes smaller than their curvilinear distance. 

Under the small gradient approximation, integrating Eq.~\eqref{eq:drag} across the filament yields
\begin{equation}\label{eq:frictional_tension}
T(x) = f(|x|-L/2)\;.
\end{equation}
where we set $\lambda=0$ under the assumption that both ends of the filament remain free during growth: i.e. $T(0)=T(L)=0$. Furthermore, with no other external forces at play, $f_{\perp}=0$ and the force driving the bending of the filament is purely tangential. Because Eqs.~\eqref{eq:small_grad} and \eqref{eq:frictional_tension} are invariant under the transformation $x\to -x$, it is sufficient to consider only one of the two halves of the filament. Focusing on the right half gives, for instance,
\begin{equation}\label{eq:right_half}
2\xi^{3}y_{xxxx}+\left(L-2x\right)y_{xx}= 0\;,\qquad 0\le x \le L/2\;,
\end{equation}
with $\xi=(B/f)^{1/3}$ a length scale. Like Eq.~\eqref{eq:euler}, this equation entails a competition between bending elasticity and compression, but the negative tension arising in response of the latter is now spatially dependent and maximal at the centre of the filament. Eq.~\eqref{eq:right_half} can be analytically solved for the curvature $\kappa=y_{xx}$ to give
\begin{equation}
\kappa(x) = \alpha \Ai\left(\frac{2x-L}{2\xi}\right) + \beta \Bi\left(\frac{2x-L}{2\xi}\right)\;,
\end{equation}
where $\Ai$ and $\Bi$ are Airy functions of the first and second kind respectively and $\alpha$ and $\beta$ integration constants determined by the boundary conditions in the range $0\le x \le L/2$. At the free end, where $x=L/2$, $\kappa(L/2)=\alpha \Ai(0)+\beta \Bi(0)=0$, but, since $\Bi(0)=\sqrt{3}\Ai(0)$, this implies $\beta=-\alpha/\sqrt{3}$. At the centre of the filament, on the other hand, the curvature is maximal, so that $\kappa_{x}(0)=0$. This requires
\begin{equation}
\Ai_{x}\left(-\frac{L}{2\xi}\right)-\frac{1}{\sqrt{3}}\,\Bi_{x}\left(-\frac{L}{2\xi}\right) = 0\;,	
\end{equation}
whose numerical solution gives the critical length
\begin{equation}\label{eq:prediction}
L_{\rm c}/\xi = 3.02981\ldots
\end{equation}
In summary, like in Euler's buckling, a negative tension arising from the lateral compression of an initially straight rod, destabilises the straight configuration against bending. On the other hand, unlike in a typical buckling problem, the tension arising from the concerted action of growth and substrate friction is not uniform across the rod, but decreases monotonically from the centre and vanishes at the ends. The deformation is, therefore, more localised than in Euler's buckling. In the next Supplementary section we will see how this peculiarity is instrumental for the formation of loops, thus for the unbinding of topological defects.

\subsection{\label{sec:looping}Loop formation}

As discussed in the main text, once a filament buckles, its growth causes an accumulation of curvature in the interior. This process continues until two or more points located at the opposite sides come into contact, thereby creating a loop at the centre of the filament. At the onset of contact, the loop is connected to the ends of the filaments by a cusp, where the tangent vector rotates by $180^{\circ}$, so that 
\begin{equation}\label{eq:contact}
x(s_{\rm c}) = x(s_{\rm c}+\ell)\;,\qquad
y(s_{\rm c}) = y(s_{\rm c}+\ell)\;,\qquad
\bm{T}(s_{\rm c})=-\bm{T}(s_{\rm c}+\ell)\;,
\end{equation}
where $s_{\rm c}$ denotes the arc-length coordinate of the point of contact and $\ell<L$ the length of the loop. As the filament continues to grow, the point of contact extends into a line. At the early stage of this process this newly formed {\em line of contact} continues to grow, thereby ``zipping-up'' the loop while this moves away from the initial point of contact. The loop, in turns, does not exhibit appreciable changes in shape and also its length appears to grow more slowly than that of the line of contact. Eventually, the loop itself buckles and its shape becomes more complex and less symmetric. 

To rationalise these observations, we go back to Eq.~\eqref{eq:elastica} describing the large deflections of a rod simultaneously subjected to a negative tension and a transverse body force. The latter, in particular, is caused by the adhesive interactions between different portions of the loop and, for simplicity, will be approximated here as a constant: i.e. $f_{\perp}={\rm const}$. Crucially, this approximation is only valid at early times, where the area of the loop is small and the opposite ends are still close to each other. Analogously, once the opposite sides of the filament come into contact, the equal and opposite reaction forces arising at the contact point mutually neutralise, thereby leaving the loop in a state of approximately uniform tension: i.e. $T(s) \approx T(s_{\rm c})=T_{\rm c}$. To make progress, we next introduce the dimensionless variables $l=s/\ell$ and $k = \kappa \ell$, so that Eq.~\eqref{eq:elastica} becomes
\begin{equation}\label{eq:similarity}
k''+\tfrac{1}{2}k^{3}-\tau k -\nu = 0\;,
\end{equation}
where the prime now indicates differentiation with respect to $l$ and we have set
\begin{equation}\label{eq:tau_nu}
\tau = \frac{T_{\rm c}\ell^{2}}{B}\;,\qquad
\nu = \frac{f_{\perp}\ell^{3}}{B}\;.
\end{equation}
Eq.~\eqref{eq:similarity} has now the structure of the classic {\em Elastica} equation, Eq.~\eqref{eq:elastica}, but with both the dimensionless tension $\tau$ and the normal load $\nu$ increasing in time with the length $\ell$ of the loop. 
 
A formal solution of this equation can be found by setting the origin of the local coordinates where the curvature is maximal or minimal, so that $k'(0)=0$. This is
\begin{equation}\label{eq:sol}
k(l) = k_{0}-\frac{1-\cn(ql,m)}{z_{1}-z_{2}\cn(ql,m)}\;,
\end{equation}
where $k(0)=k_{0}$ is the curvature at the origin and $\cn(\cdot,\cdot)$ Jacobi's elliptic cosine defined from the inverse function of the incomplete elliptic integral of the first kind. That is
\begin{equation}
u = F(\phi,m) = \int_{0}^{\phi}\frac{\D{}t}{\sqrt{1-m^{2}\sin^{2} t}}\;,
\end{equation}
from which $\phi=\am(u,m)$ is the Jacobi amplitude and $\cn(u,m)=\cos\phi$. The parameters $z_{1}$ and $z_{2}$ correspond to the solutions of the quadratic equation 
\begin{equation}
z^{2}-2\alpha z + 2\alpha\beta -(\beta^{2}+\gamma^{2}) = 0\;,
\end{equation}
where $\alpha$, $\beta$ and $\gamma$ are themselves related to the roots a cubic polynomial via the factorization
\begin{equation}\label{eq:roots}
y^{3}+\frac{Q}{P}\,y^{2}+\frac{k_{0}}{P}\,y-\frac{1}{4P} 
= (y-\alpha)[(y-\beta)^{2}+\gamma^{2}]\;,
\end{equation}
where we have set
\begin{equation}
P = k_{0}^{3}-2\left(\tau k_{0}+\nu\right)\;,\qquad
Q = \tau-\frac{3}{2}\,k_{0}^{2}\;.
\end{equation}
Assuming $y_{1}=\alpha\in\mathbb{R}$, Eq.~\eqref{eq:roots} allows expressing the remaining two roots of the polynomial on the left-hand side of the equation as $y_{2}=\beta+i
\gamma$, $y_{3}=\beta-i\gamma$. Assuming three real roots would have instead led to a contradiction~\cite{veerapaneni_analytical_2009}. Finally, the parameters $q$ and $m$ in Eq.~\eqref{eq:sol} are given by
\begin{equation}
q = \sqrt{\frac{P(z_{1}-z_{2})}{2}}\;,\qquad
m = \sqrt{\frac{\beta-z_{2}}{z_{1}-z_{2}}}\;.
\end{equation}
A complete derivation of this solution and other details about Eq.~\eqref{eq:similarity} can be found in~\cite{giomi_softly_2013} and references therein; here, we limit ourselves to discussing the implications of Eq.~\eqref{eq:sol} to the formation and early times stability of loops. To this end, we first notice that, as soon as the opposite end of the rod comes into contact, symmetry requires that $(\bm{R}-\bm{R}_{0})\cdot\bm{N}=0$ at the point of contact, hence 
\begin{equation}\label{eq:curvature_at_contact}
k(l_{\rm c}) = -\sqrt{\frac{2T}{B}}\;,
\end{equation}
by virtue of Eq.~\eqref{eq:R_vs_F} and having set $l_{\rm c}=s_{\rm c}/\ell$. Furthermore, as the curvature is minimal at the contact point, $k'(l_{\rm c})=0$ and $\bm{F}(l_{\rm c})=\bm{0}$. As the filament continues to grow, on the other hand, the normal load experienced by the loop increases until -- when $\nu$ is larger than a critical value $\nu_{0}$ -- the initially isolated point of contact extends into a line, along which $k(l_{\rm c})=k'(l_{\rm c})=0$. By virtue of Eqs.~\eqref{eq:tau_nu} and \eqref{eq:curvature_at_contact}, the latter implies $\tau=0$ uniformly across the loop.

\begin{figure}
\centering
\includegraphics[width=\textwidth]
{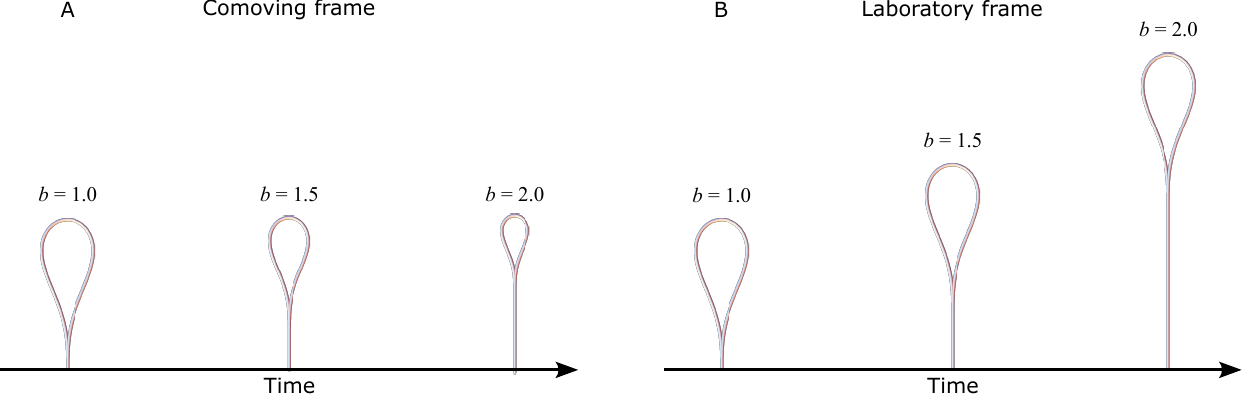}
\caption{\label{fig:loop_formation}Examples of the post-buckling dynamics at the early stage of loop formation. (A) Configurations of the loop in the comoving frame, where the arc-length coordinate $l=s/\ell$ spans the fixed range $0\le l \le 1$. As the filament grows, the scale factor $b$ increases in time, thus effectively enhancing the magnitude of the normal load $\nu$. In this frame, the latter implies a shrinkage of the filament. (B) In the laboratory frame, the aforementioned effect is compensated by the growth of the filament, thereby preserving both the shape and size of the loop. The strong adhesive forces acting along the line of contact ``seal'' the $+1/2$ defect encircled by the loop, thus preventing this from annihilating shortly after buckling.}
\end{figure}

To reconstruct the shape of the loop at the onset of this process, it is convenient to set the origin of the rescaled arc-length coordinate at the mid-point of the loop, where the curvature is maximal, and calculate the constants $k_{0}$ and $\nu_{0}$ from the equations
\begin{equation}\label{eq:constraints}
k(l_{\rm c}) = 0\;,\qquad
\int_{-l_{\rm c}}^{l_{\rm c}}\D{}l\,k = \pi\;,
\end{equation}
where $l_{\rm c}=1/2$. Using Eq.~\eqref{eq:sol} with $\tau=0$ and solving \eqref{eq:constraints} with respect to $k_{0}$ and $\nu_{0}$ gives
\begin{equation}
k_{0} = 14.4082\ldots,\qquad 
\nu_{0} = 640.964\ldots
\end{equation}
Remarkably, further growth does move the loop apart from the point of contact, but without altering its shape and size. To illustrate this concept, we notice that Eq.~\eqref{eq:similarity} is invariant with respect to a special transformation consisting of a rescaling of the loop -- i.e. $l\to l/b$, with $b>1$ a scale factor -- and a renormalization of the material parameters~\cite{flaherty_postbuck_1972,djondjorov_analytic_2011}. That is
\begin{equation}
\left\{l,\,k,\,\tau,\,\nu\right\} \to 
\left\{\frac{l}{b},\,bk,\,b^{2}\tau,\,b^{3}\nu\right\}\;.
\end{equation}
As a consequence, the shape of the loop obtained for $\nu>\nu_{0}$ is {\em similar} to that found for $\nu=\nu_{0}$ up a scaling factor $b=(\nu/\nu_{0})^{1/3}>1$. When parametrized in terms of the rescaled arc-length coordinate $l$, this implies a shrinkage of the loop, with $l_{\rm c} = \pm 1/(2b)$ marking the new location of the point of contact, and a simultaneous extension of the line of contact, whose length is now $\ell_{\rm line}/\ell=1-2l_{\rm c}$ (see Fig.~\ref{fig:loop_formation}A). More interestingly, restoring the original arc-length coordinate $ -\ell/2 \le s \le \ell/2$, with $\ell=\ell(t)$ the growing length of the loop, we see that $\ell_{\rm c} = \pm \ell/(2b) = \pm \ell_{0}/2$, with $\ell_{0}$ being the same length of the loop when $\nu=\nu_{0}$ (see Fig.~~\ref{fig:loop_formation}B). As anticipated, this implies that both the shape and size of the loop remain unaltered while moving apart from the initial point of contact. We notice that the appearance of a line of contact has the important effect of ``locking'' the loop as it extends, thereby protecting the newly unbound pair of $\pm 1/2$ nematic disclinations from an immediate annihilation. 

\newpage
\section{Bead spring numerical model}
We model the bacterial colonies using a two-dimensional bead spring model, similarly to \cite{isele2015self,kurjahn2024collective}. The filaments are described as a series of beads connected to each other via stiff elastic springs, where each bead represents the centre of a bacterium. These also interact with each other via exclusion and attraction forces, through a modified Lennard-Jones potential. The springs expand over time to simulate filament growth, and new beads are added to represent cell division. To account for interactions with the substrate, we also add a constant friction force to oppose motion, and a viscous drag. The evolution of the position $\boldsymbol{r}_i$ of the $i$-th bead, assuming an overdamped regime, is thus given by:
\begin{equation}
\boldsymbol{\nu}_i \cdot \Dot{\boldsymbol{r}_i} = -\nabla_i U^{\left(el\right)} - \nabla_i U^{\left(bend\right)} - \nabla_i U^{\left(int\right)} - \boldsymbol{F}_i^{\left(fric\right)}, 
\end{equation}
The dynamics is purely deterministic: we neglect noise effects in the growth or in the substrate friction in order to focus on the interplay between the different forces. 
We will now give a more precise explanation of the model and the forces involved.
\subsection{Interactions}
Our filaments are driven by compression, bending and friction forces. Regarding the first one, we assume nonlinear elasticity, considering a fourth-order potential between consecutive beads, leading to a cubic force:
\begin{equation}   U^{\left(el\right)}_{i,i+1} =\frac{k_e}{4l_{i}^3}\left(|\boldsymbol{r}_{i,i+1}|-l_{i}\right)^4 \quad \Rightarrow \quad \boldsymbol{F}^{\left(el\right)}_{i,i+1}=\frac{k_e}{l_{i}^3}(|\boldsymbol{r}_{i,i+1}|-l_{i})^3  \frac{\boldsymbol{r}_{i,i+1}}{|\boldsymbol{r}_{i,i+1}|},
\end{equation}
%
where $\boldsymbol{r}_{i,i+1}=\boldsymbol{r}_{i+1}-\boldsymbol{r}_i$ is the vector connecting two consecutive beads, $l_i$ is the natural length of the spring connecting $i$ and $i+1$ and $k_e$ is the elastic constant.

Bond flexibility is governed by a finite-angle generalisation of a standard harmonic potential: 
\begin{equation}\label{eq:bending_pot}
    U^{\left(bend\right)}_i=-k'_b \log \left[\cos\left(\frac{\phi_i}{\phi_{max}}\frac{\pi}{2}\right)\right],
\end{equation}

where $k'_b$ is the bending stiffness of the filament and $\phi_i$ is the angle formed by the two vectors $\boldsymbol{r}_{i,i+1}$ and $\boldsymbol{r}_{i-1,i}$. 
The resulting torque is linear for small $\phi$, but it saturates for $\phi=\phi_{max}$, preventing both unphysical deformations and numerical instabilities:
\begin{equation}
    M_i=k_b \tan\left(\frac{\phi_i}{\phi_{max}}\frac{\pi}{2}\right),
\end{equation}
where we have reabsorbed the extra constants that come out of the logarithm into $k_b$. This torque is equivalent to a triplet of forces applied on the $i$-th bead and its neighbours:
\begin{gather}      \boldsymbol{F}^{\left(bend\right)}_{i-1,i}=\frac{M_i}{|\boldsymbol{r}_{i-1,i}|+|\boldsymbol{r}_{i,i+1}| }\hat{r}_{i-1,i}^{\perp}, \qquad        \boldsymbol{F}^{\left(bend\right)}_{i+1,i}=\frac{M_i}{|\boldsymbol{r}_{i-1,i}|+|\boldsymbol{r}_{i,i+1}|} \hat{r}_{i,i+1}^{\perp}, \nonumber \\[1.2ex]
\boldsymbol{F}^{\left(bend\right)}_{i,i}=-\boldsymbol{F}^{\left(bend\right)}_{i-1,i}-\boldsymbol{F}^{\left(bend\right)}_{i+1,i},
\end{gather}
where $\hat{r}_{ij}^{\perp}$ is a unitary vector perpendicular to $\boldsymbol{r}_{ij}$ such that it points in the direction of the restoring force (Diagram 1).
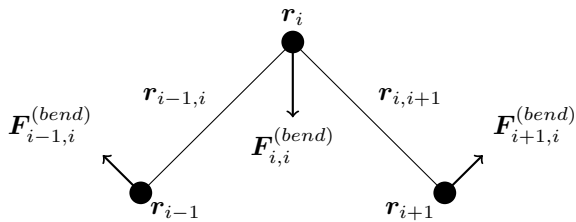
\begin{figure}[h!]
	\centering
	\begin{tikzpicture}
	
	\draw (0,0) -- (2,2)node[pos=0.5,anchor=south east] {$\boldsymbol{r}_{i-1,i}$};
	\draw (4,0) -- (2,2)node[pos=0.5,anchor=south west] {$\boldsymbol{r}_{i,i+1}$};
	\fill (0,0) circle (0.15cm)node[anchor=north west] {$\boldsymbol{r}_{i-1}$};
	\fill (2,2) circle (0.15cm)node[above=3pt] {$\boldsymbol{r}_i$};
	\fill (4,0) circle (0.15cm)node[anchor=north east] {$\boldsymbol{r}_{i+1}$};
	\draw[thick,->] (0,0) -- (-0.5,0.5)node[anchor=south east] {$\boldsymbol{F}^{\left(bend\right)}_{i-1,i}$};
	\draw[thick,->] (2,2) -- (2,1)node[anchor=north] {$\boldsymbol{F}^{\left(bend\right)}_{i,i}$};
	\draw[thick,->] (4,0) -- (4.5,0.5)node[anchor=south west] {$\boldsymbol{F}^{\left(bend\right)}_{i+1,i}$};
	\end{tikzpicture}
	
	\caption{\textbf{Diagram 1}: Diagram of the forces associated to bending stiffness.}
	\label{fig:bending}
\end{figure}

To take in account both steric and adhesion effects, non-consecutive beads (belonging to the same or to different filaments) interact with each other via a  modified Lennard Jones potential $U^{\left( int \right)}$. 
\begin{equation}
U_{LJ}(d_{ij})= 4\epsilon \left[ \left(\frac{\sigma}{d_{ij}}\right)^{12}-\left(\frac{\sigma}{d_{ij}}\right)^6 \right]
\end{equation}
\[   
U^{\left( int \right)}(d_{ij})=
\begin{cases}
U_{LJ}(d_{ij})-U_{LJ}(d_\text{max}) &\quad\text{if }d_{ij}<d_\text{max}\\
0 &\quad\text{if }d_{ij}\geq d_\text{max}\\
\end{cases}
\]

where $d_{ij}$ is the distance between the $i$-th particle and its projection on $\boldsymbol{r}_{j,j+1}$ (Diagram 2).
This corresponds to a short-range exclusion force between filaments for $d_{ij}<\sigma$, preventing overlapping, and to a midrange attraction for $\sigma<d_{ij}<d_\text{max}$, recreating the stickiness of the bacteria.\\
\\
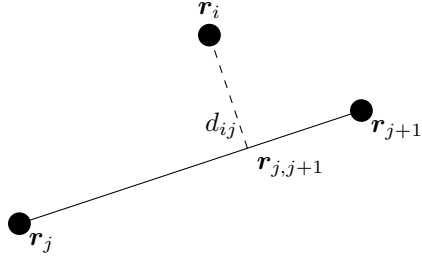
\begin{figure}[h!]
	\centering
	\begin{tikzpicture}
	\draw (2,2) -- (2,2)node[pos=0.5,anchor=south east] {};
	\draw (-0.5,-0.5) -- (4,1)node[pos=0.5,anchor=south west] {};
	
	\fill (2,2) circle (0.15cm)node[above = 3pt] {$\boldsymbol{r}_i$};
	\fill (-0.5,-0.5) circle (0.15cm)node[anchor=north west] {$\boldsymbol{r}_j$};
	
	\fill (4,1) circle (0.15cm)node[anchor=north west] {$\boldsymbol{r}_{j+1}$};
	\draw[dashed] (2,2) -- (2.5,0.5)node[anchor=south east] {$d_{ij}$};
	\draw (2.5,0.5) node [anchor = north west] {$\boldsymbol{r}_{j,j+1}$};
	\end{tikzpicture}
		
	\caption{\textbf{Diagram 2:} Diagram of the distance between the $i$-th particle and its projection on $\boldsymbol{r}_{j,j+1}$}
	\label{fig:interaction}
\end{figure}

Following \cite{yaman_emergence_2019}, interactions with the environment give rise to two opposing-motion effects: a  uniform friction and a viscous drag. The former is modelled as $\boldsymbol{F}^{\left(fric\right)}_{i}=-\mu \hat{\boldsymbol{v}}_i$, where $\hat{\boldsymbol{v}}_i=\dot{\boldsymbol{r}}_i|/\dot{\boldsymbol{r}}_i|$ denotes the direction of motion of the $i$-th bead, and $\mu$ is the friction coefficient. The latter, taking in account the elongated shape of the filaments, is expressed through a viscous drag tensor $\boldsymbol{\nu}_i$,
\begin{equation}
\boldsymbol{\nu}_i=\nu_\parallel \boldsymbol{t}_i\otimes \boldsymbol{t}_i+\nu_\perp\left(\boldsymbol{I}-\boldsymbol{t}_i\otimes \boldsymbol{t}_i\right)
\end{equation}
where $\boldsymbol{I}$ is the identity tensor, $\boldsymbol{t}_i=\left({\boldsymbol{r}_{i+1}-\boldsymbol{r}_{i-1}} \right)/\left|\boldsymbol{r}_{i+1}-\boldsymbol{r}_{i-1}\right|$ is the tangent vector of the filament at the $i$-th bead, and $\nu_\parallel$ and $\nu_\perp$ are the drag coefficients for motion in the direction, respectively, parallel and perpendicular to $\boldsymbol{t}_i$.

Neglecting thermal noise and self-propulsion, the complete equation of motion is therefore:
\begin{equation}
\boldsymbol{\nu}_i \cdot\dot{\boldsymbol{r}}_i=\boldsymbol{F}^{\left( el \right)}_{i\ i+1}+\boldsymbol{F}^{\left( el \right)}_{i\ i-1}+\boldsymbol{F}^{\left( bend \right)}_{i,i-1}+\boldsymbol{F}^{\left( bend \right)}_{i,i}+\boldsymbol{F}^{\left( bend \right)}_{i,i+1}+\boldsymbol{F}^{\left( fric \right)}_{i}+\sum_{j\neq i}\boldsymbol{F}^{\left( int \right)}_{ij}
\end{equation}

\subsection{Filament growth}
The activity in this system stems from the growth of the filaments, which we model by an increase of the natural length of the springs connecting nodes with a growth rate $\lambda$, as done in \cite{yaman_emergence_2019}.
When the distance between adjacent nodes reaches a critical value $l_c$, a new node having their average speed is inserted in the midpoint.

To prevent unlimited overcrowding, the effective growth of the natural length $l_i$ of a single node is governed by a generalised logistic law, depending on the spring compression $|\boldsymbol{r}_{i,i+1}|/l_i$:
\begin{equation}
\frac{dl_{i}}{dt}=\lambda_0 \left[1+\left(\frac{\alpha l_{i}}{|\boldsymbol{r}_{i,i+1}|}\right)^n\right]^{-1}.
\end{equation}
A free filament, not subject to a relevant compression, is thus characterised, at the level of individual nodes, by a constant growth rate $\lambda \approx \lambda_0$, reflecting in an exponential growth of the total length. Crowding, which generates an increase in mechanical stresses, slows down the local expansion of the filament until the critical spring length $|\boldsymbol{r}_{i,i+1}|=\alpha l_i$ is exceeded. The parameter $\alpha<1$ sets the compression at which the growth rate is reduced by a factor $2$, while the steepness of this decrease is dictated by the value of the exponent $n$.
\\
\subsection{Simulations}
The simulations are carried out in a square periodic domain of size $L\times L$. Initial conditions are generated by placing $N \times N$ filaments, having initial natural length $l_{i0}$, within a central square subdomain of size $(2/5 L) \times (2/5 L)$, leaving free space for colony growth. As in the experiments, we considered both parallel, perpendicular and random initial orientations of the filaments. Instability is triggered by superimposing to the flat initial profile a weak localised Gaussian-shaped perturbation.
A small positional and angular noise is added to the initial condition, to introduce variability in the evolution of different filaments.

Temporal evolution is implemented by using a standard 4-th order Runge-Kutta scheme. Similarly to the experiments, during their evolution, filaments expand also in the surrounding empty area, but we analyse only the bulk of the system, to avoid border effects. The values of the parameters used are given in Table \ref{tab:parameters}.

We use the number of filaments $N=\{4,8,16,18,20,22\}$ at fixed system size $L$ as control parameter. For each value of $N$ we average over three different realisations of the initial noise.

\begin{table}[h!]
	\centering
	\begin{tabular}{|c|c||c|c|}
		\hline
		\multicolumn{4}{|c|}{Numerical parameters} \\ \hline
		$k_e$  & 10.0 & $\nu_\parallel$ & 0.3   \\ \hline
		$k_b$ & 85.0 & $\nu_\perp$  & 0.45 \\ \hline
		$\phi_\text{max}$ & $\pi/4$ &$\mu$  & 0.6 \\ \hline
		$\epsilon$  & 2.0 &$\lambda_0$  & 0.1 \\ \hline
		$\sigma$  & 0.1 & $n$  & 18\\ \hline
		$d_\text{max}$  & 0.14 &$\alpha$  & 2/9 \\ \hline
		$l_c$  & 0.05 & $l_{i,0}$ & $0.03$ \\ \hline
		$L$  & 64.0 & & \\ \hline
	\end{tabular}
	\caption{Values of the numerical parameters used in the multiple filaments simulations, as described in the text.}
	\label{tab:parameters}
\end{table}

\subsection{Comparison with experiments}

We rescale the spatial and temporal units of the simulations using, respectively, the buckling critical length and the filament doubling time as reference scales.

The buckling critical length is obtained from single-filament simulations initialised with a small perturbation at the filament centre. We define the buckling length as the filament length at which the local slope first reaches 60°. Averaging over different realisations yields $l_b = 2.76 \pm 0.06$ in simulation units, corresponding to $l_b= \left( 87.6 \pm \ 1.9 \right) \mu m$ in experiments.

The reference timescale is the time required for a filament to double its length. In simulations this corresponds to $T= l_c /(2\lambda_0) = 0.25$. Using the experimentally measured growth rate $\mu$ (Sec. \ref{subsec1.1}), this maps to $T= \ln{2}/\mu = 24$ min.

\subsection{Comparison with theory}
The empirical potential \eqref{eq:bending_pot} can be related to the bending term in Eq.~\eqref{eq:small_grad} assuming $\phi_i \ll \phi_{max}$ and constant separation length $|\boldsymbol{r}_{i-1,i}| \approx|\boldsymbol{r}_{i,i+1}| \equiv \Delta s$:
\begin{align}
U^{\left(bend\right)}\approx \sum_{i=1}^{N-1} k'_b\left(\frac{\pi}{2\phi_{max}}\right)^2\phi_i^2=\sum_{i=1}^{N-1}\frac{k_b\pi}{2\phi_{max}}\phi_i^2\equiv \sum_{i=1}^{N-1}K|\boldsymbol{r}_{i-1,i}|^2\phi_i^2 = \nonumber \\
= 2K\sum_{i=1}^{N-1}\left[|\boldsymbol{r}_{i-1,i}|^2-|\boldsymbol{r}_{i-1,i}|^2\left(1-\frac{1}{2}\phi_i^2\right)\right]
 \approx K\sum_{i=1}^{N-1}\left[ |\boldsymbol{r}_{i-1,i}|^2+|\boldsymbol{r}_{i,i+1}|^2-2|\boldsymbol{r}_{i-1,i}||\boldsymbol{r}_{i,i+1}|\cos(\phi_i)\right]= \nonumber \\ = K\sum_{i=1}^{N-1}\left(\boldsymbol{r}_{i-1,i}-\boldsymbol{r}_{i,i+1}\right)^2.
\end{align}
The associated force is the gradient of the bending energy with respect to position:
\begin{equation}
\begin{split}
    \boldsymbol{F}_i^{\left(bend\right)}=\boldsymbol{\nabla}_{r_i}U^{\left(bend\right)} &=\boldsymbol{\nabla}_{r_i}K\sum_{j=1}^{N-1}\left(\boldsymbol{r}_{j-1,j}-\boldsymbol{r}_{j,j+1}\right)^2=\boldsymbol{\nabla}_{r_i}K\sum_{j=1}^{N-1}\left(\boldsymbol{r}_{j-1}-2\boldsymbol{r}_{j}+\boldsymbol{r}_{j+1}\right)^2\\
&=2K\left(\boldsymbol{r}_{i-2}-4\boldsymbol{r}_{i-1}+6\boldsymbol{r}_{i}-4\boldsymbol{r}_{i+1}+\boldsymbol{r}_{i+2}\right),
\end{split}
\end{equation}
which in the continuum limit can be rewritten in terms of the fourth derivative of $\boldsymbol{r}$ with respect to the arclength:
\begin{equation}
    \boldsymbol{F}^{\left(bend\right)} \left( \boldsymbol{r} \right)=2K\Delta s^4 \frac{\partial^4 \boldsymbol{r} }{\partial s^4}.
\end{equation}
The constant $B$ in Eq.~\eqref{eq:small_grad}) is therefore connected to the parameters of our model as:
\begin{equation}
    B=2K\Delta s^4=\frac{k_b\pi\Delta s^2}{\phi_{max}},
\end{equation}
thus the length scale $\xi$ is given by:
\begin{equation}
    \xi\equiv \left(\frac{B}{\mu}\right)^{1/3}=\left(\frac{k_b\pi\Delta s^2}{\mu\phi_{max}}\right)^{1/3}.
\end{equation}
Considering an average separation $\Delta s\approx0.038$, from the values in Tab.~\ref{tab:parameters} we obtain $\xi\approx0.927$ in simulation units. Considering $L_c = l_b$ we find the ratio  $L_c/\xi=2.97 \pm 0.07$, fully consistent with the value $\approx 3.03$ predicted by the stability analysis (Eq.~\eqref{eq:prediction}).

\newpage

\section{Supplementary figures}\label{supsec2}

\figuresection{Figure S1 - Scheme of Capillarity-Assisted Particle Assembly set-up}\label{subsec2.1}

\begin{figure}[H]
\centering
\makebox[\textwidth][c]{
    \includegraphics[width=\textwidth]{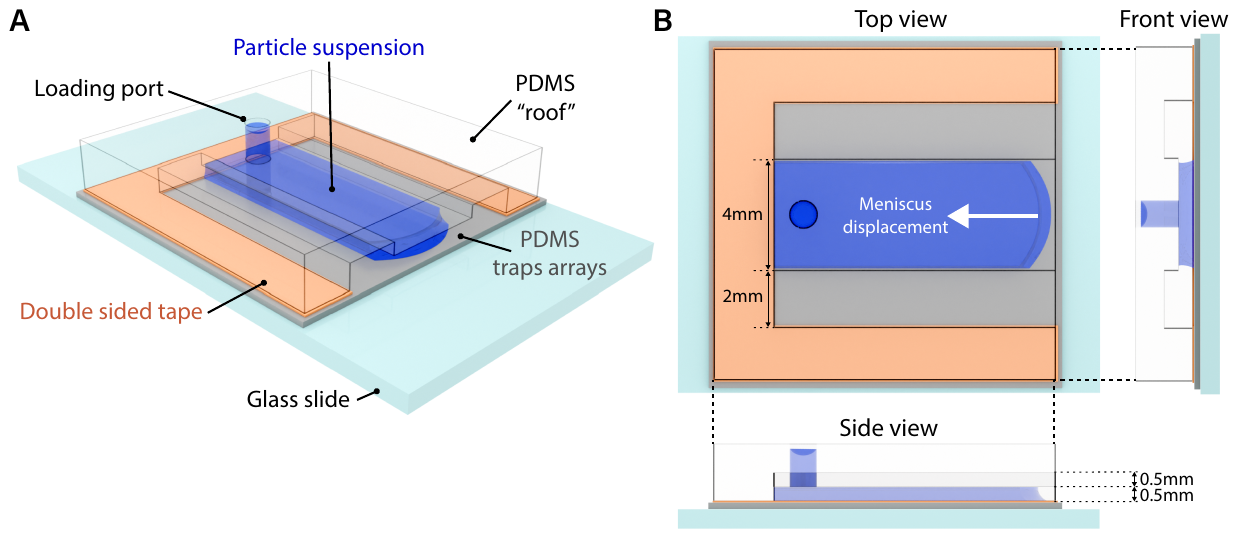}
    }
\caption{(A) 3D view of the set-up assembly. (B) 2D view of the top, front and side.}\label{figsup_CAPARoof}
\end{figure}

\figuresection{Figure S2 - Bacterial growth rate and germination time distributions}\label{subsec2.2}

\begin{figure}[H]
\centering
\makebox[\textwidth][c]{
    \includegraphics[width=\textwidth]{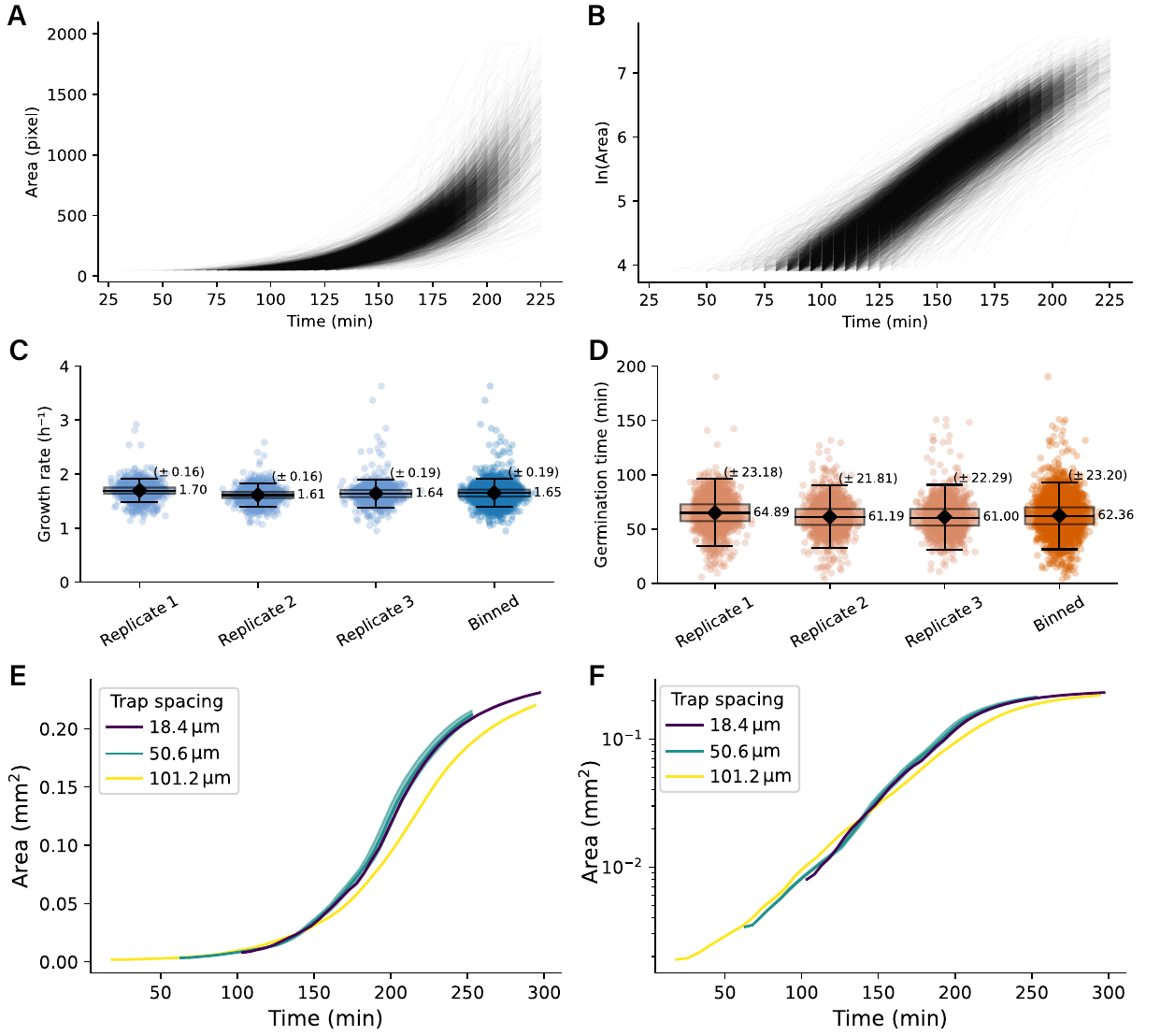}
    }
\caption{(A) Time evolution of bacteria-segmented area with exponential growth. (B) The corresponding natural logarithm transform of the growth curves used to fit linear regressions. (C) Area growth rate and (D) germination time distribution among replicate experiments. (E) Numerical time evolution of bacteria-filled area with exponential growth. (F) The corresponding natural logarithm transform of the growth curves.}\label{figsup_GR-Germ}
\end{figure}

\figuresection{Figure S3 - Topological defect velocities}\label{subsec2.3}

\begin{figure}[H]
\centering
\makebox[\textwidth][c]{
    \includegraphics[width=\textwidth]{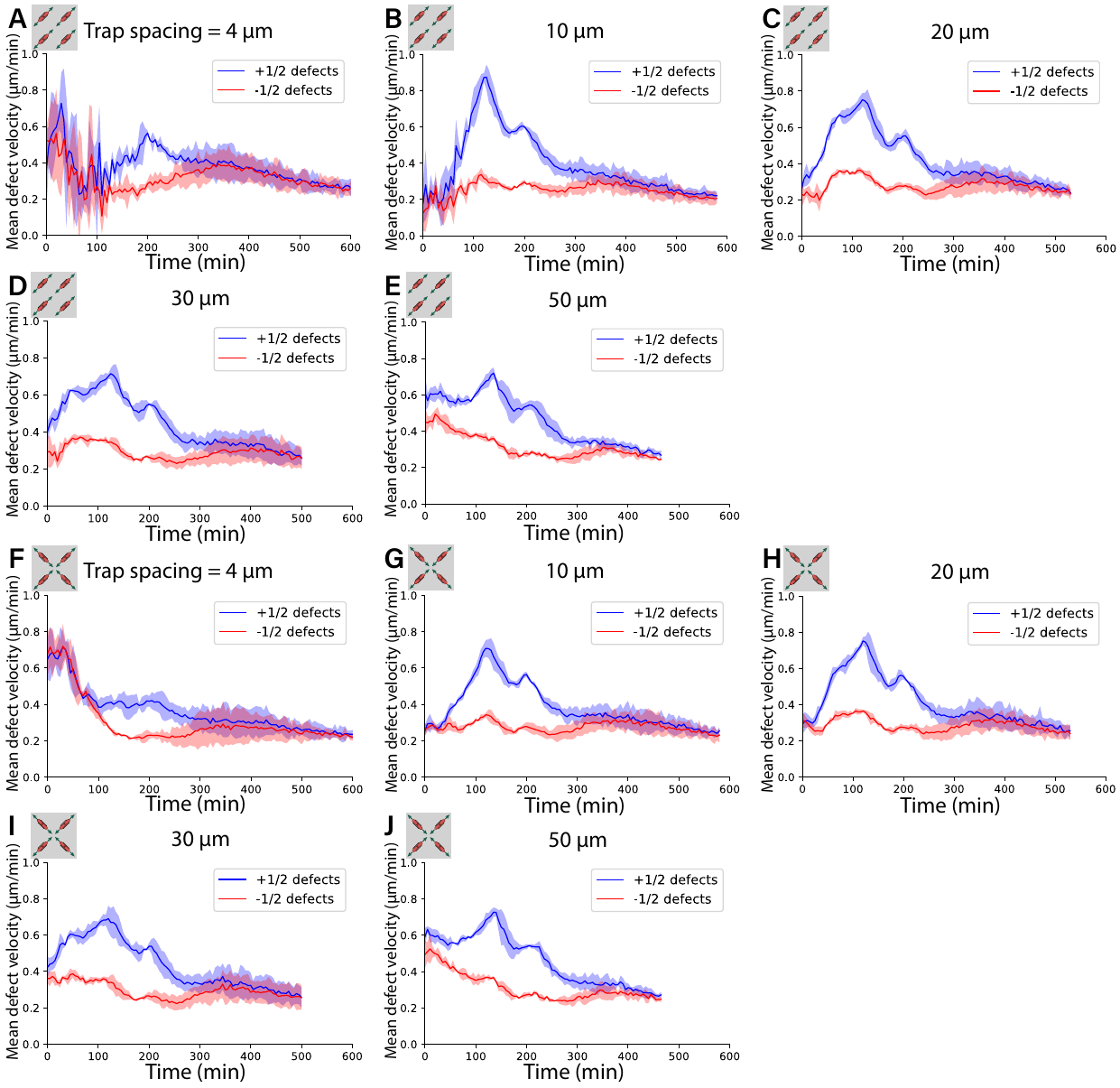}
    }
\caption{(A-E) Time evolution of topological defect velocity for +1/2 and -1/2 defects for parallel orientation at different trap spacings. (F-J) Time evolution of topological defect velocity for +1/2 and -1/2 defects for perpendicular orientation at different trap spacings. Solid lines represent mean values and shaded areas represent ±one standard deviation for three technical replicates.}\label{figsup_DefVelo}
\end{figure}

\figuresection{Figure S4 - PIV cell flow velocities as proxi for growth}\label{subsec2.4}

\begin{figure}[H]
\centering
\makebox[\textwidth][c]{
    \includegraphics[width=\textwidth]{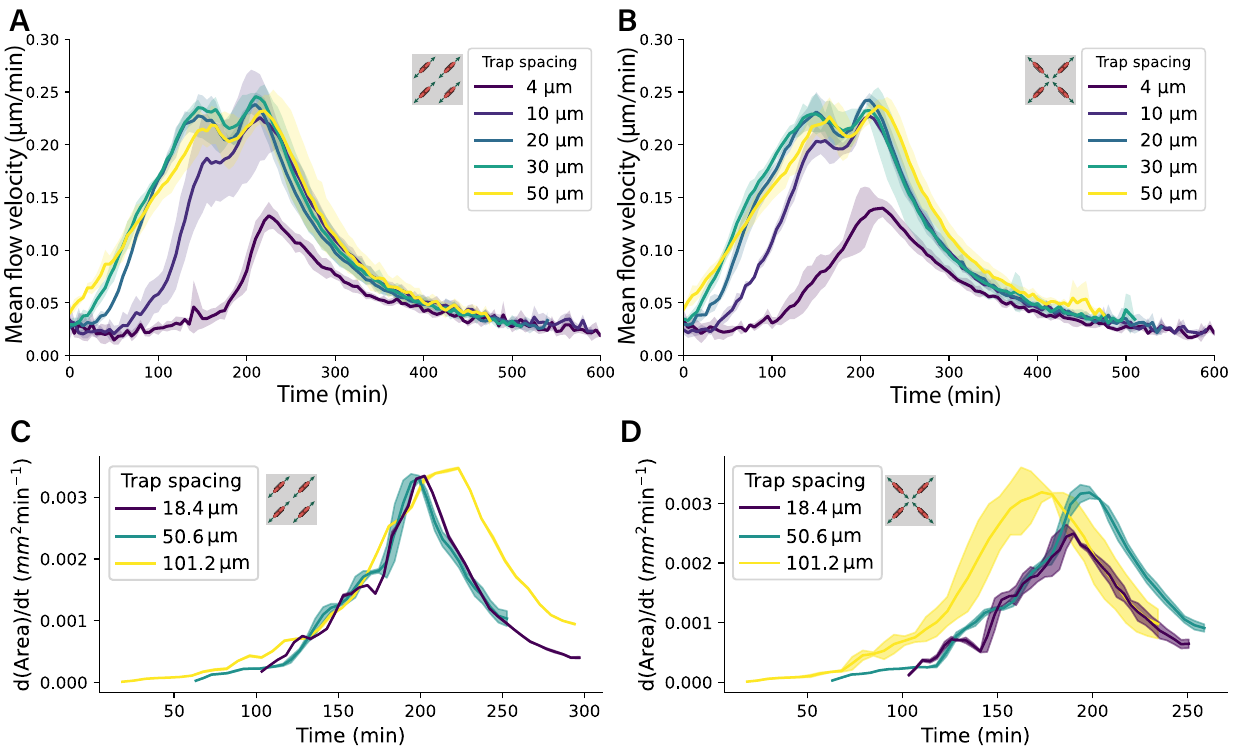}
    }
\caption{(A) Time evolution of cell flow estimation for parallel orientation at different trap spacings. (B) Time evolution of cell flow estimation for perpendicular orientation at different trap spacings. Solid lines represent mean values and shaded areas represent ±one standard deviation for three technical replicates. (C) Numerical time evolution of growth rate for the parallel orientation at different trap spacings. (D) Numerical time evolution of growth rate for the perpendicular orientation at different trap spacings. Solid lines represent mean values and shaded areas represent ±one standard deviation for three technical replicates.}\label{figsup_PIVvelo}
\end{figure}

\figuresection{Figure S5 - Nematic and -1/2 defects orientations for parallel, perpendicular and random orientations for bioreplicates}\label{subsec2.5}

\begin{figure}[H]
\centering
\makebox[\textwidth][c]{
    \includegraphics[width=\textwidth]{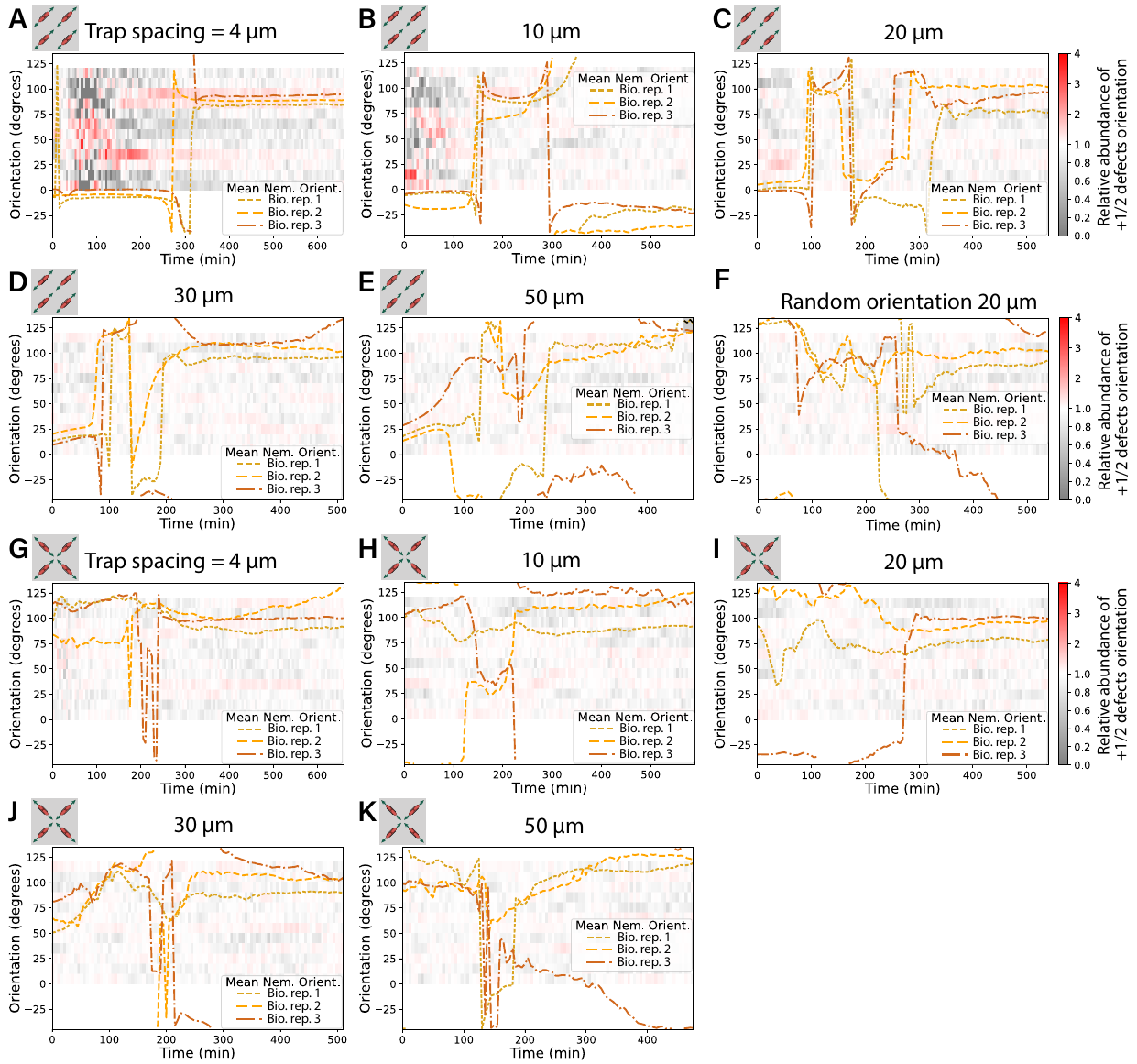}
    }
\caption{(A-E) Time evolution of the mean nematic orientation (dashed lines, bio-replicate separate) and -1/2 topological defect orientation (heat map for relative abundance to uniform distribution, bio-replicate binned) for parallel orientation at different trap spacings, (F) for random orientation at 20~µm trap spacing, and (G-K) for perpendicular orientation at different trap spacings. Grey corresponds to depletion and red corresponds to enrichment of defect orientation compared to the uniform distribution, represented as white. $-90^{\circ}$ is equivalent to $30^{\circ}$ for $-1/2$ defects because of their 3-fold symmetry }\label{figsup_OrienBioRepNEG}
\end{figure}

\figuresection{Figure S6 - Nematic and +1/2 defects orientations for parallel, perpendicular and random orientations for bioreplicates}\label{subsec2.6}

\begin{figure}[H]
\centering
\makebox[\textwidth][c]{
    \includegraphics[width=\textwidth]{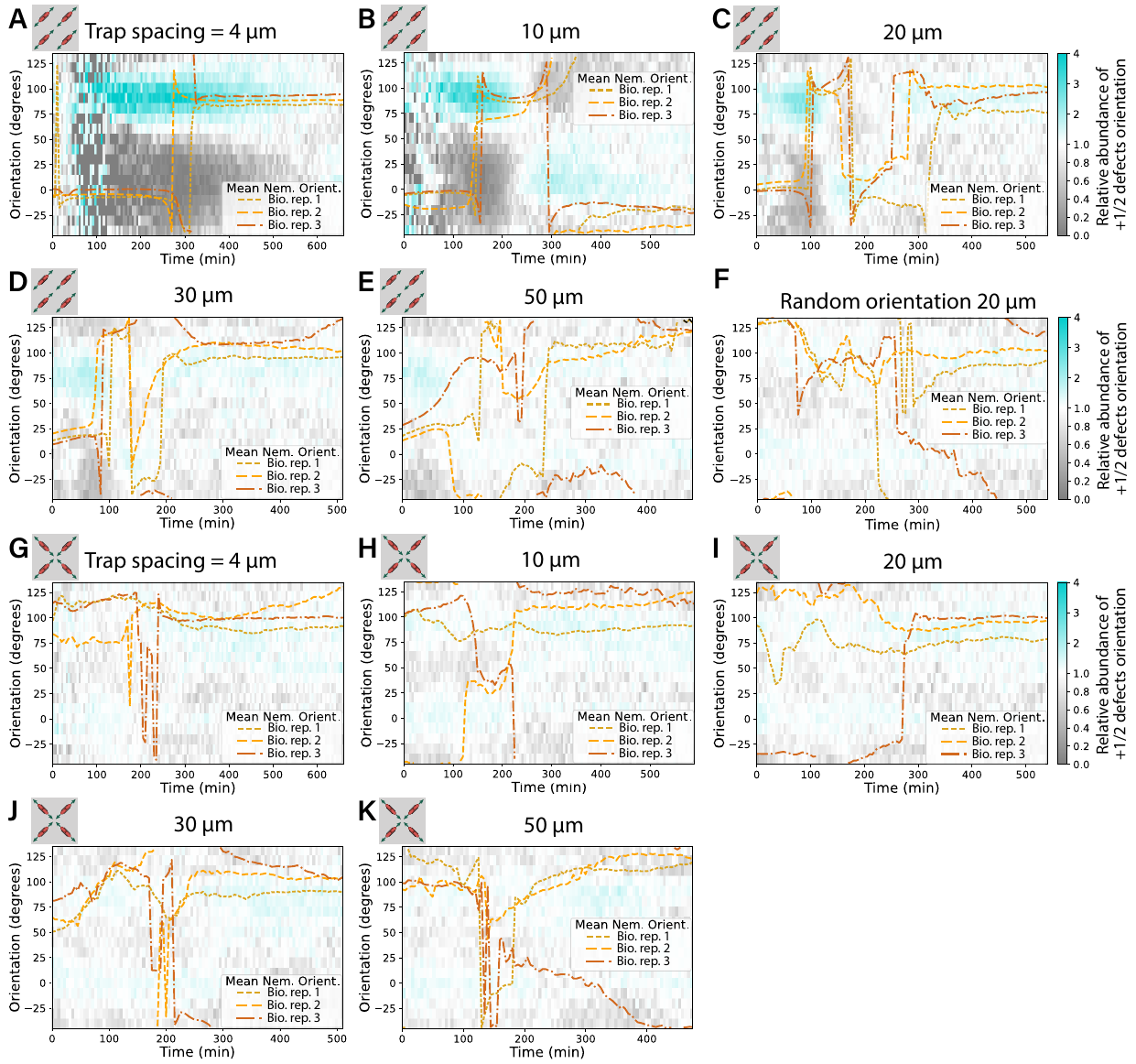}
    }
\caption{(A-E) Time evolution of the mean nematic orientation (dashed lines, bio-replicate separate) and +1/2 topological defect orientation (heat map for relative abundance to uniform distribution, bio-replicate binned) for parallel orientation at different trap spacings, (F) for random orientation at 20~µm trap spacing, and (G-K) for perpendicular orientation at different trap spacings. Grey corresponds to depletion and cyan corresponds to enrichment of defect orientation compared to the uniform distribution, represented as white. }\label{figsup_OrienBioRepPOS}
\end{figure}

\figuresection{Figure S7 - Cell thickness and cell-cell distance corresponds to grating spacing}\label{subsec2.7}

\begin{figure}[H]
\centering
\makebox[\textwidth][c]{
    \includegraphics[width=\textwidth]{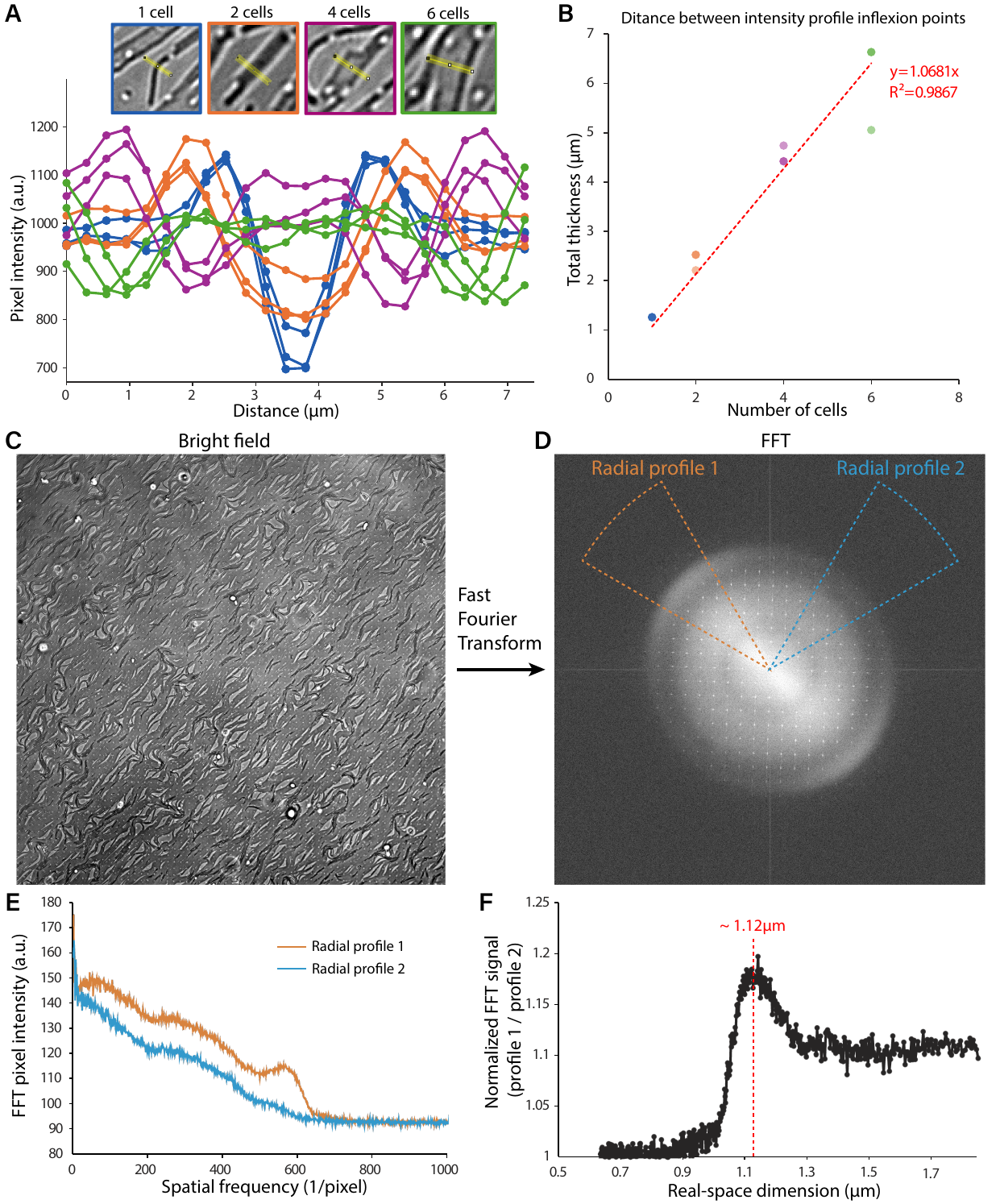}
    }
\caption{(A) Intensity profiles of cell clusters of 1, 2, 4 and 6 cell thickness from bright-field images, inset represent one example of each. (B) Measure distance between intensity profile inflection points. Linear regression fit give a cell thickness of 1.07~µm. (C) Example of bright field image of bacteria filament with high nematic order oriented diagonally. (D) Corresponding fast Fourier transform image. Two radial profiles are measured along the bacteria short axis (profile 1) and long axis (profile 2). (E) Radial intensity profiles of the FFT image. (F) Normalised FFT radial intensity profile 1 against profile 2. The peak correspond to the cell thickness feature only present in profile 1, giving a cell thickness of 1.12~µm.}\label{figsup_CellSpacing}
\end{figure}

\figuresection{Figure S8 - Polarised light microscopy and structural colour snapshot of 2D patterns with local nematic orientation control}\label{subsec2.8}

\begin{figure}[H]
\centering
\makebox[\textwidth][c]{
    \includegraphics[width=\textwidth]{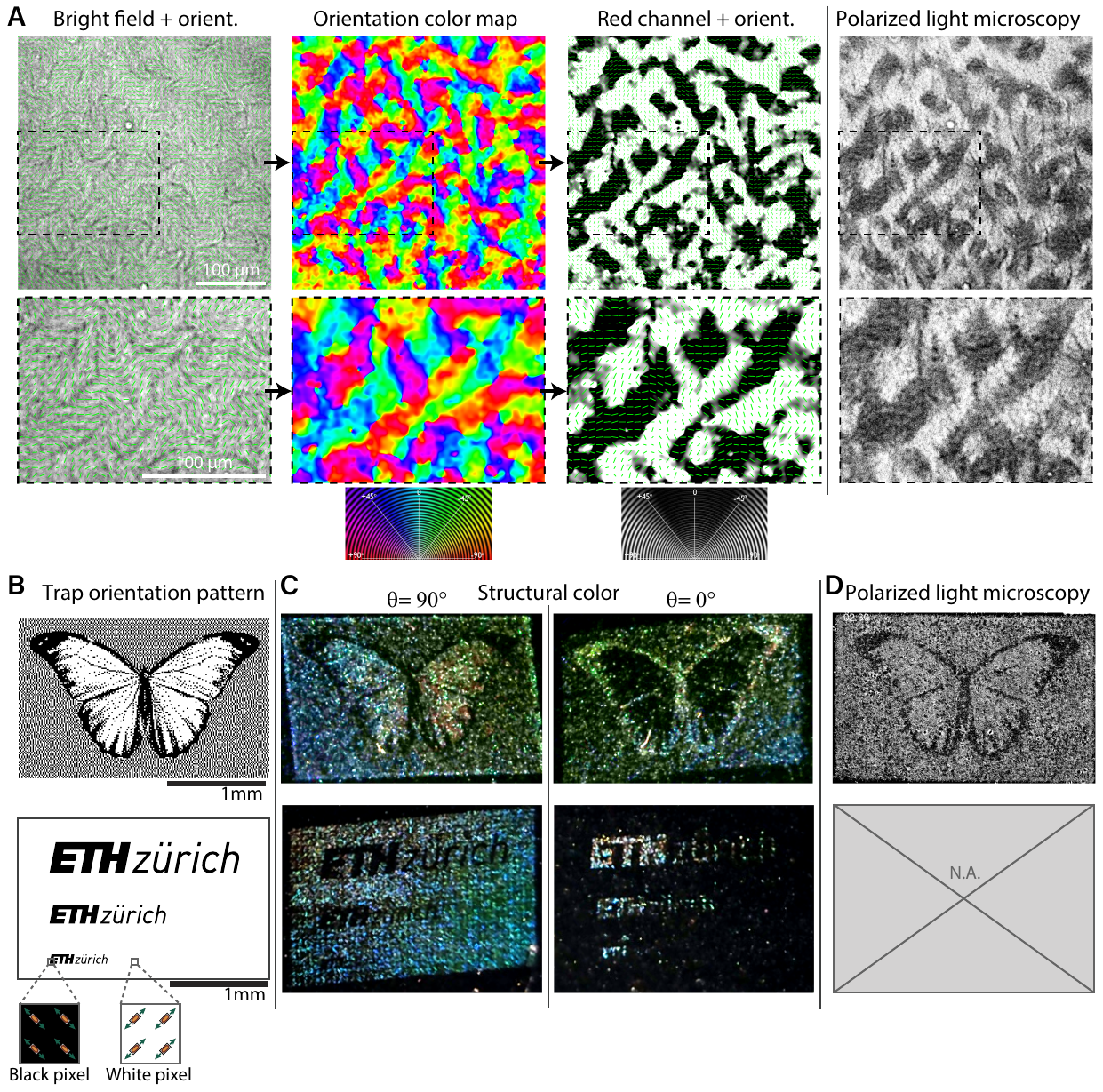}
    }
\caption{(A) Bright field image (first column) of a bacteria 2D film in active nematic phase with the filament orientation field superimposed in green. The orientation field can be represented as a colour map (second column) with the hue value representing the orientations between +90 and -90 degrees (colour scale below the image). The red channel of such orientation colour map (third column) gives an quasi binary image, with bright regions for vertical orientations (90$\pm$45 degrees) and dark region for horizontal orientations (0$\pm$45 degrees). The image obtained with polarised light microscopy (fourth column) is highly similar to such red channel image, with light extinction localised where the filaments are oriented orthogonally to the polariser axis. The $\pi$-periodic angular extinction differs from the classic $\pi/2$-periodicity arising from homogeneous linear birefringence. We speculate that images are instead dominated by polarisation-dependent diffraction from the oriented bacterial filaments. Polarised light microscopy makes readily apparent the local nematic orientation within the bacterial film by contrasting the two primary orthogonal directions. (B) Binary image of the 2D trap orientation pattern with black and white pixel oriented orthogonally. (C) Structural colour images of the bacteria film growing from such trap pattern, imaged at confluence but before buckling at two orthogonal imaging orientations. (D) Polarised light microscopy image of such 2D patterns.}\label{figsup_POM-Color}
\end{figure}

\newpage
\section{Supplementary Movies}\label{supsec3}
\figuresection{Supplementary Movie 1}\label{sec3.1}
Supplementary Movie 1 (AVI, 21.1 MB): Isolated chain buckling

\figuresection{Supplementary Movie 2}\label{sec3.2}
Supplementary Movie 2 (AVI, 316 MB): Bacterial chains grown from 4~µm spaced spore arrays with perpendicular or parallel orientations with topological defects superimposed. +1/2  defects are represented as cyan arrow and -1/2 ones as red trefoils. Only defects tracked for at least 3 frames are considered valid and are superimposed.

\figuresection{Supplementary Movie 3}\label{sec3.3}
Supplementary Movie 3 (AVI, 268 MB): Bacterial chains grown from 10~µm spaced spore arrays with perpendicular or parallel orientations with topological defects superimposed. +1/2  defects are represented as cyan arrow and -1/2 ones as red trefoils. Only defects tracked for at least 3 frames are considered valid and are superimposed.

\figuresection{Supplementary Movie 4}\label{sec3.4}
Supplementary Movie 4 (AVI, 255 MB): Bacterial chains grown from 20~µm spaced spore arrays with perpendicular or parallel orientations with topological defects superimposed. +1/2  defects are represented as cyan arrow and -1/2 ones as red trefoils. Only defects tracked for at least 3 frames are considered valid and are superimposed.

\figuresection{Supplementary Movie 5}\label{sec3.5}
Supplementary Movie 5 (AVI, 253 MB): Bacterial chains grown from 30~µm spaced spore arrays with perpendicular or parallel orientations with topological defects superimposed. +1/2  defects are represented as cyan arrow and -1/2 ones as red trefoils. Only defects tracked for at least 3 frames are considered valid and are superimposed.

\figuresection{Supplementary Movie 6}\label{sec3.6}
Supplementary Movie 6 (AVI, 233 MB): Bacterial chains grown from 50~µm spaced spore arrays with perpendicular or parallel orientations with topological defects superimposed. +1/2  defects are represented as cyan arrow and -1/2 ones as red trefoils. Only defects tracked for at least 3 frames are considered valid and are superimposed.

\figuresection{Supplementary Movie 7}\label{sec3.7}
Supplementary Movie 7 (AVI, 2.60 MB): Living diffraction grating growing from 4~µm spaced spore arrays with parallel orientations imaged at $\alpha$~=~30° and at different angle $\theta$~=~90°, 45°, 0°, -45° every 30min. 

(D) Patterning living diffraction grating spatially by varying trap orientations. A binary image of a butterfly is used as template, black pixels and white pixels correspond to a SO-NE and a NO-SE orientation respectively (90° difference). When imaged at the quasi-perfect nematic order state ($\sim$7~h of growth), the image template is observed as structural colouration pattern at $\theta$~=~90° for the positive and $\theta$~=~0° for the negative version of the template.

\figuresection{Supplementary Movie 8}\label{sec3.8}
Supplementary Movie 8 (AVI, 1.55 MB): Binary living diffraction grating image of a butterfly imaged at $\alpha$~=~30° and at $\theta$~=~90° for the positive and $\theta$~=~0° for the negative version of the template.

\newpage

\bibliography{arxiv}

\makeatletter
\immediate\write\@auxout{%
  \string\newlabel{dummy-label-for-aux}{{\string\relax}{1}}%
}
\makeatother

\end{document}